\documentclass[prb,reprint,aps,floatfix,showpacs,superscriptaddress,longbibliography]{revtex4-1}
\usepackage{amssymb}
\usepackage{bbm}
\usepackage{dsfont}

\usepackage{amsmath}
\usepackage{graphicx}
\usepackage[caption=false]{subfig}
\usepackage[colorlinks=true,linkcolor=blue,anchorcolor=red,citecolor=blue,urlcolor=blue]{hyperref}
\begin{document}
\title{Interaction effects in Graphene in a weak magnetic field}
\author{Ke Wang}
\email{kewang@umass.edu}
\affiliation{%
	Department of Physics, University of Massachusetts, Amherst, MA 01003, USA}
\author{M. E. Raikh}
\affiliation{Department of Physics and Astronomy, University of Utah, Salt Lake City, UT 84112, USA}
\author{T. A. Sedrakyan}%
\email{tsedrakyan@umass.edu}
\affiliation{%
	Department of Physics, University of Massachusetts, Amherst, MA 01003, USA}

\date{\today}
\begin{abstract}
	A weak perpendicular magnetic field, $B$,
	breaks the chiral symmetry of each valley in the electron spectrum of graphene, preserving the overall chiral symmetry in the Brillouin zone.
	We explore the consequences of this symmetry breaking
	for the interaction effects in graphene. In particular,
	we demonstrate that the electron-electron interaction lifetime acquires
	an anomalous $B$-dependence.
	Also, the ballistic zero-bias anomaly, $\delta \nu(\omega)$, where $\omega$ is the
	energy measured from the Fermi level,
	emerges at a weak $B$ and has the form $\delta\nu(B)\sim B^2/\omega^2$.
	Temperature dependence of the magnetic-field corrections to the thermodynamic characteristics of graphene is also anomalous.
	We discuss experimental manifestations of the effects predicted.
	The microscopic origin of the $B$-field sensitivity is an extra phase acquired
	by the electron wave-function resulting from the chirality-induced pseudospin precession.

\end{abstract}
\maketitle

{\em Introduction.} Electron spectrum in graphene possesses a
chiral (pseudo-spin) structure\cite{prl84Semenoff,Katsnelson2006}.
Two pseudospin projections are identified with two points, $K$ and $K'$,
of the Brillouin zone near which the spectrum is characterized
by a massless Dirac dispersion.
Numerous consequences of the Dirac spectrum of graphene
for the disorder and interaction
effects were established,
see e.g.
Refs. \onlinecite{prl06Altland,prl06Aleiner,prl08Tse,rmp09Castroneto,rmp11Dassarma,rmp12Kotov,IOP12Gordon,np12Nandkishore,nature19Dutreix,arxiv20Agarwal,prb19maiti,prx20Pack,prl21Leeb,prl21Sbierski,prlGuo21,prl21Bouaziz,prb21Rostami,prb21Narozhny}.

One distinctive feature of the graphene bandstructure is the absence of backscattering from the impurities.
This feature is a consequence of orthogonality of the spinors corresponding to the wave vectors ${\bf k}$
and $-{\bf k}$. In turn, the absence of backscattering leads to the suppression of the oscillations
of electron density (Friedel oscillations\cite{52Friedel}) created by an impurity in graphene.\cite{prb99Raikh,prl06Cheianov}
In the ballistic regime\cite{prb97Glazman,zala1,zala2,zala3}, electron scattering  from individual impurities dressed by
the Friedel oscillations is responsible for a zero-bias anomaly $\propto \ln \omega$
in conventional 2D electron gas. Here $\omega$ is the energy measured from the Fermi level
and the condition $\omega\tau \gg 1$, where $\tau$ is the elastic scattering time, is implied.
Fast decay of the Friedel oscillations suggests that zero-bias anomaly in graphene
is absent.\cite{prl07Macdonald,prl06Cheianov}
More detailed study\cite{prb07Glazman} indicated that it is the Hartree
correction which is absent in graphene, while the Fock correction, originating
from the forward scattering, is still present.

%$K$ and $K'$ points of the Brillouin zone (BZ), leading to a variety of remarkable phenomena in electron transport and physics of correlations\cite{prl06Altland,prl06Aleiner,prl08Tse,rmp09Castroneto,rmp11Dassarma,rmp12Kotov,IOP12Gordon,np12Nandkishore,nature19Dutreix,arxiv20Agarwal,prb19maiti,prx20Pack,prl21Leeb,prl21Sbierski,prlGuo21,prl21Bouaziz,prb21Rostami,prb21Narozhny}.

%The chirality of electrons suppresses their back-scattering process. One manifestation of the suppression property is that Friedel oscillations (FO) of electron density around an impurity in graphene decay much faster\cite{prb99Raikh,prl06Cheianov} than in 2D electron gas\cite{52Friedel}.  Due to the faster-decaying FO, the logarithmic zero-bias anomaly in graphene is absent for the Coulomb
%impurities\cite{prl07Macdonald,prl06Cheianov} and only emerges for the atomically sharp impurities in the ballistic regime ($ \omega \tau >1$)~\cite{prb07Glazman}. Here $\omega$ is the energy measured from the Fermi level and $\tau$ is the scattering time. % Upon using the expression of $M_0$, one obtains that the leading terms in $H$ vanishes. This reflects that the chiral symmetry suppress the back-scattering and also indicates the vanishing of the zero-bias anomaly that originates from Friedel oscillations. The Fock diagram is different. With the atomically sharp impurities, the leading term in $F$ does {\em not} vanish. This in fact reproduces the result in Ref.~\onlinecite{prb07Glazman} that the zero-bias emerges from forward scatterings in Dirac materials }

In the absence of impurities, electron-electron interactions in 2D electron gas
%In the Fermi-liquid theory in two dimensions (2D), electron-electron interactions
cause non-analytic corrections\cite{prb82Quinn,prb96Macdonald, prb96Sarma, prb03Chubukov}
to the self-energy, $\Sigma(\omega)$.
%as a function of energy.
At low temperatures, $T\ll \omega$,
the imaginary part of  self-energy has the form
$\tau_{ee}(\omega)^{-1} \sim (\omega^2/E_F ) \ln(E_F/\omega)$, where $E_F$
is the Fermi energy. Correspondingly, the real part of self-energy behaves as
$\text{Re}\Sigma(\omega) \sim \omega^2\text{sgn}(\omega)$.
At finite $T$, interactions cause a correction to the specific heat\cite{jpsj99Misawa,prl93Coffey}
$\delta C(T) \propto T^2$. Microscopically, the above corrections emerge in the random-phase
approximation. Their derivation is so general that it is natural to expect that, in doped graphene, the
interaction corrections have the same Fermi-liquid form.\cite{prb07das}

%$\tau$, of quasi-particles. At zero temperature, the scattering rate of quasi-particles as a function of energy obeys a logarithmic behavior, $\tau^{-1}(\omega)\propto (\omega^2/E_F ) \ln(E_F/\omega)$, where $E_F$ is the Fermi energy.
%
%Similarly, the correction to the real part of the self energy is non-analytic, $\text{Re}\Sigma(\omega) \propto \omega^2\text{sgn}(\omega)$. The analogous consideration for thermodynamic characteristics gives the quadratic in temperature correction to the specific heat\cite{jpsj99Misawa,prl93Coffey}, $\delta C(T) \propto T^2$.

%The common understanding of the interactions in the doped graphene is that the forward-scattering process leads to observable characteristics in good agreement with the 2D Fermi-liquid (FL) theory\cite{prb07das}. It is natural to expect non-analytic interaction corrections to observables in the Fermi-liquid description of the doped graphene.   %The results above have been obtained both within random phase approximation (RPA) scheme and perturbation theory and are in qualitative agreement with each other.   %In this context, one remarkable result is the zero-bias anomaly of the tunneling DOS in the ballistic regime in the presence of impurity scatterings in the graphene monolayer\cite{prb07Glazman}.

In the present letter, we identify the interaction effects specific to graphene.
These effects emerge in the presence of a weak magnetic field. Their
origin is the field-induced lifting of chiral symmetry in $K$ and $K^\prime$ valleys of graphene while preserving the overall symmetry.
To capture these effects, one should go beyond the random-phase approximation.

%we report the emergence of novel interaction-induced effects in transport and
%thermodynamic characteristics of the doped graphene monolayer in the presence of a weak
%non-quantizing magnetic field.% that are beyond the conventional RPA scheme.

With regard to ballistic zero-bias anomaly, lifting of the chiral symmetry in the field, $B$,
gives rise to the contribution $\propto B^2/\omega^2$, which can be even {\em stronger} than the
zero-field contribution.\cite{prb07Glazman}
A formal difference between the calculations of the ballistic
zero-bias anomaly in electron gas with parabolic spectrum and in graphene
is that the Green functions, which enter into the calculation, have a
matrix  structure in graphene. Without this matrix structure,
the $B$-sensitive contributions to the tunnel conductance {\em cancel out}.

A natural energy scale imposed by the field, $B$, in graphene is $\omega_0 =v_F/R_L$,
where $R_L\propto B^{-1}$ is the Larmour radius. Quantization of the energy levels
can be neglected for $\omega \gg \omega_0$. We show that the $B$-dependent correction
to the thermodynamic characteristics of the clean graphene can be conveniently expressed
in terms of $\omega_0$. Namely, the corrections to the imaginary and real parts of
self-energy behave as $\text{Im}\left[\Sigma(\omega,B)-\Sigma(\omega,0)\right]
\sim \omega_0^2 E_F^{-1}\ln(\omega/T)$ and
$\text{Re}\left[\Sigma(\omega,B)-\Sigma(\omega,0)\right]\sim \omega_0^2 E_F^{-1} \text{sgn} (\omega)$,
respectively. On the basis of these results we draw the consequences for observables.
Namely, we show that the $B$-dependent correction to the specific heat is temperature-independent in
a  wide temperature interval.

{\em Electrons in a weak magnetic field.}
%Here we discuss propagators of Dirac electrons which leads to singularities in field-dependent interaction-induced observables discussed in the introduction.
The Hamiltonian of monolayer graphene which incorporates the $B$ field in the Landau gauge reads
\begin{eqnarray}
\hat{H}_{B}= v_F \left[ (p_x-eBy) \hat{\Sigma}_x + p_y \hat{\Sigma}_y \right].
\end{eqnarray}
Here $v_F$ is the Fermi velocity.  Here $\hat{r}= \mathbf{r} /r$, $ {\boldsymbol \Sigma}=(\Sigma_x, \Sigma_y)$ and $\Sigma_x= \hat{\tau}_z \otimes \hat{\sigma}_x$, $\Sigma_y= \hat{\tau}_z \otimes \hat{\sigma}_y$. The Pauli matrices $\hat\sigma_{i}$ act in the space of $A$ and $B$ sublattices of the honeycomb lattice and $\hat{\tau}$ is the Pauli matrix distinguishing between two Dirac points in Graphene. Diagonalizing the Hamiltonian, one finds that the linear spectrum is transformed into a non-uniform ladders of spectrum, $\sqrt{2 n} v_F/l$. Here $n\geq 0$ and $l=\sqrt{\hbar/eB}$ is the magnetic length. Under a weak field, the spectrum around the Fermi level, $E_F$, can be linearized as
%is approximated by
$\sqrt{2 n} v_F/l \simeq E_F+ (n-N_F) v_F (k_F l^2)^{-1}$, where $N_F=(k_F l)^2/2$. This yields the expression for the
effective cyclotron frequency $\omega_0= v_F(k_F l^2)^{-1}$.

%This in turn introduces an effective cyclotron-frequency $\omega_0= v_F(k_F l^2)^{-1}$. This cyclotron-frequency is the %fundamental energy scale in this paper.

The Feynman propagator of free Dirac electrons is known to possess a non-trivial matrix structure. Namely, in the absence of
%a
magnetic field, the propagator in the real space is given by~\cite{prb07Glazman}   \begin{eqnarray}
\label{green}
G_\omega(\mathbf{r})=\frac{k_F}{2v_F} \sqrt{\frac{1}{2k_Fr}} e^{i\text{sgn}(\omega) \Phi_0(r)}    M_0,
\end{eqnarray}
where the phase $\Phi_0(r)=k_Fr+{ \omega r}/v_F + \pi/4$ and the matrix $ M_0$ is given by $   M_0=\left(\text{sgn}(\omega)+i (2k_Fr)^{-1}\right) \hat{r}\cdot \mathbf{\Sigma}+\hat{I}$.  Here $\hat{r}= \mathbf{r} /r$, $ {\boldsymbol \Sigma}=(\Sigma_x, \Sigma_y)$ and $\hat{I}$ is the identity matrix. This matrix structure reflects the chiral symmetry of electrons: fast
decay of the Friedel oscillations\cite{prl06Cheianov} and the absence of a zero-bias anomaly\cite{prb07Glazman} are the consequences of this matrix form.

%and leads to many interesting properties in graphene, including faster decaying of Friedel oscillations and absence of zero-bias anomaly\cite{prb07Glazman}.

Presence of magnetic field modifies the gauge-invariant part of the electron propagator by breaking the chiral symmetry of the electrons in the vicinity of the Dirac point.
%The
Field-induced modification of Eq.~\ref{green}
amounts to the changes of $\Phi_0(r)$ and $M_0(r)$. The phase $\Phi_0$ becomes $\Phi=\Phi_0-  r^3/(24k_Fl^4)$, which is due to the curving of the semiclassical trajectory\cite{prl07Sedrakyan,prl08Sedrakyan,prl08Tigran} in a weak field. In graphene, due to the matrix structure of  the Hamiltonian,  the identity matrix, $\hat{I}$, in $M_0$
%becomes
transforms into a new 4-dimensional field-dependent matrix.
This matrix contains $\hat\Sigma_z$,
%which depends on $\hat\Sigma_z$
%as a result, it
and thus, does not commute with
$M_0$. This is because $M_0$
%which
contains the matrices $\hat\Sigma_{x,y}$.
%dependent  part of the propagator.
Here $\hat{\Sigma}_{z}=\hat{\sigma}_{z} \otimes \hat{\tau}_0$,
where $\tau_0$ is
%the
a $2\times2$ unit matrix.

Specific form of the matrix, $M$, is the following
%Namely, $M_0$ becomes the weak-field dependent matrix $M$, given by
\begin{align}
\label{3}
M(\mathbf{r},\text{sgn}(\omega))\simeq  M_0 -i\text{sgn}(\omega)\varphi(r)\hat{\Sigma}_z
-\frac{\varphi(r)^2}{2}\hat{I},
\end{align}
where $\varphi(r)=\omega_0 r/(2v_F)$ is half of the angle corresponding to the arc of the Larmour circle with length $r$.
Eq.~(\ref{3}) applies in the domain $k^{-1}_F<r<k_F l^2=R_L$.
%{\color{red} where $k^{-1}_F<r<k_F l^2$.

\begin{figure}
	\centering
	\includegraphics[scale=0.35]{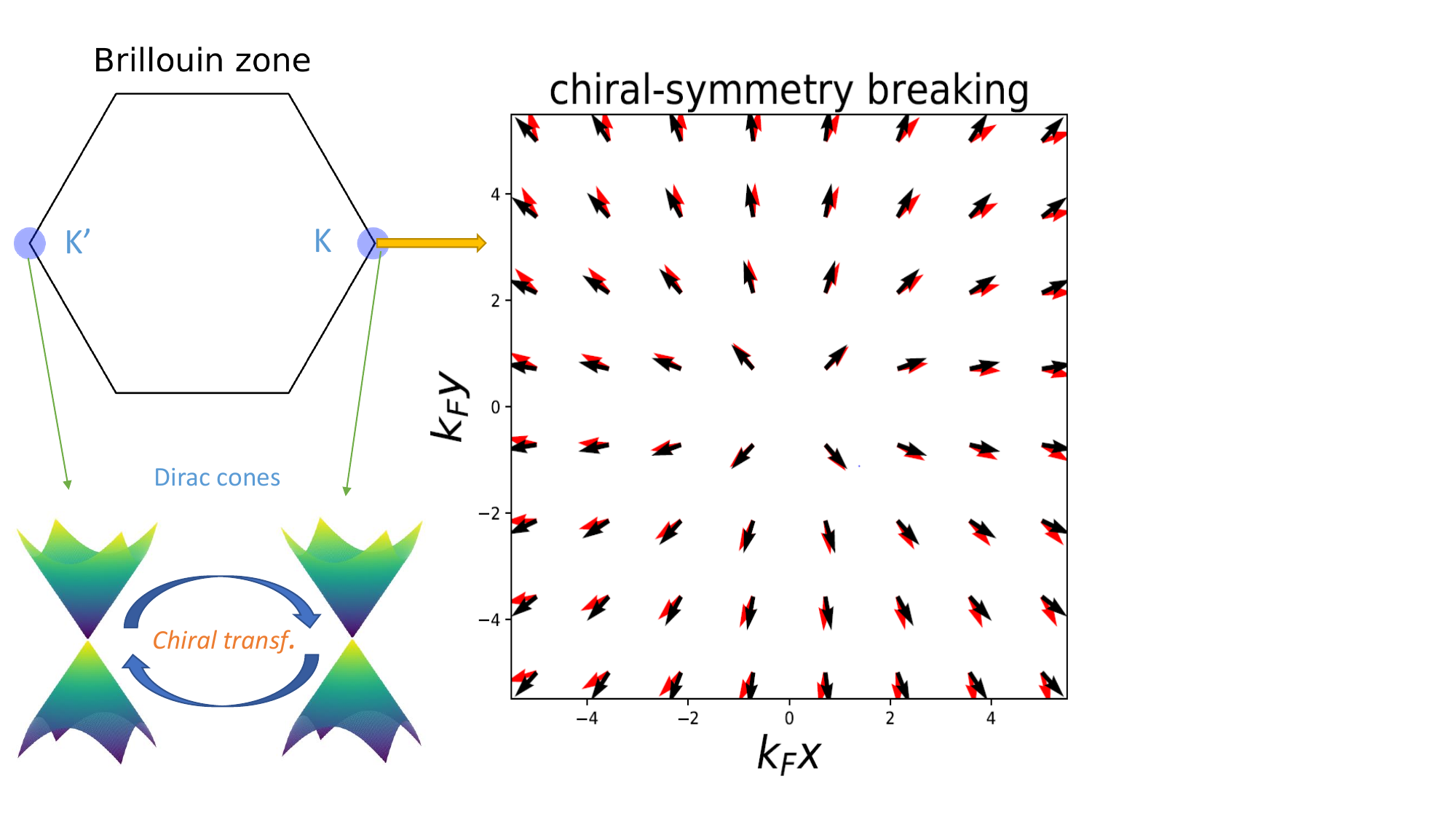}
	\caption{ (Color online) The left panel depicts the Brillouin zone of graphene. Around the $K$ and $K'$ valleys, the spectrum is Dirac-like, supporting low-energy Hamiltonians $\hat{H}_{K}$ and $\hat{H}_{K'}$ that are connected via a chiral transformation, $H_K=\hat{\sigma}_z H_{K'} \hat{\sigma}_z$.
		The right panel depicts the vector field $\mathbf{v}_K(\mathbf{r})$ (for definition, see the footnote\cite{footnote1}), at $K$-valley. The dark (black) vectors field represents the $\mathbf{v}_K(\mathbf{r})$ at zero magnetic field. The grey (red) vector field is the $\mathbf{v}_K(\mathbf{r})$ at a weak but non-zero magnetic field. Here we take $\omega_0 /(2E_F)=0.07$. The figure shows the chiral symmetry of the state at $B=0$. At finite $B$, the chiral-symmetry in one valley is broken. In the leading approximation, the angle between two vector fields is proportional to $\varphi(r)$.  Importantly, the chiral transformation leads to the relation, $\mathbf{v}_{K'}(\mathbf{r})=\mathbf{v}_{K}(-\mathbf{r})$, manifesting the chiral symmetry of the whole system.
	}
	
	%Fock Digram. It gives the leading magnetic corrections to quasiparticle lifetimes. (d) Hartree diagram. It is the %first diagram of RPA. }
	\label{pictorial}
\end{figure}

The {\em pseudospin} structure of the term $\sim\varphi(r)\hat{\Sigma}_z$ in the propagator, while preserving the chiral symmetry of the system\cite{prl84Semenoff,prb20wang}, reflects the field-induced breaking thereof around a single Dirac cone.
see Fig.~\ref{pictorial} for a graphical representation of this effect.

%due to the non-commutativity of $\hat{\Sigma}_z$ and $\hat{\Sigma}_{x,y}$. The $\varphi(r)$-related terms are responsible %for the sensitivity for the magnetic field\cite{prb20wang}. The derivation of Eq.~\ref{3} is presented in %Ref.~\onlinecite{appendix}.}

%This form of the propagator and the origin of the magnetic phase has been discussed in our earlier publication Ref.~\onlinecite{prb20wang}, where it is shown that the weak magnetic field enhances the backscattering in graphene, leading to a novel, non-decaying form of Friedel oscillations.

%In the perturbation theory in interactions, the leading Feynman diagrams in 2D that describe the renormalization of the Green's function and describe the  FL properties and the tunneling DOS are the Hartree and Fock diagrams shown in Fig.~\ref{diagram1}.

\begin{figure}
	\centering
	\includegraphics[scale=0.45]{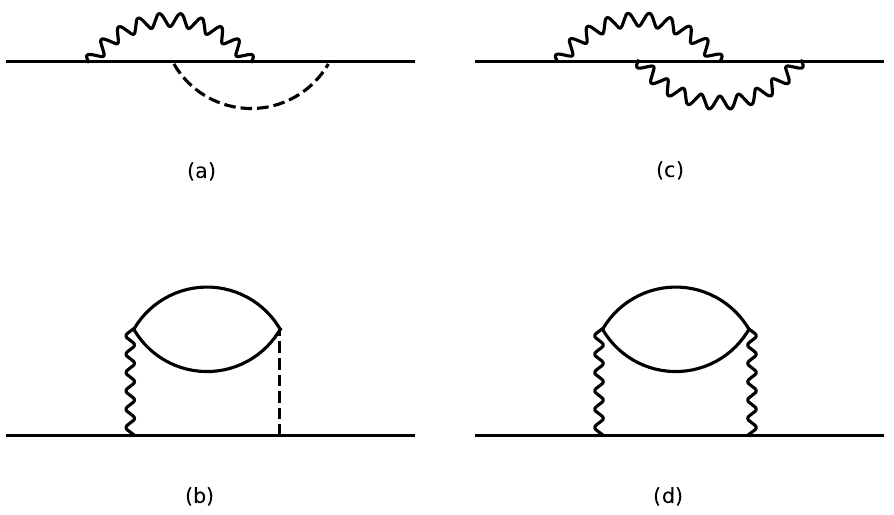}
	\caption{Diagrams for the corrections to the Green function.
		Solid lines represent the Feynman propagators. Wavy lines represent the electron-electron interactions. (a) represents the Fock diagram involving
		a single-impurity scattering. It yields a leading contribution to the
		$B^2$ ballistic zero-bias anomaly. (b) represents a Hartree diagram involving a single-impurity scattering. It is insensitive to a weak magnetic field.
		(c) and (d) represent, respectively, the Fock and Hartree diagrams for the $B^2$ correction
		to the electron lifetime. Unlike the Hartee diagram, which is the first
		diagram of the RPA sequence, diagram
		(c) yields an anomalous temperature dependence.}
	%Fock Digram. It gives the leading magnetic corrections to quasiparticle lifetimes. (d) Hartree diagram. It is the %first diagram of RPA. }
	\label{diagram1}
\end{figure}

In general, the ballistic correction to the density of states is given by  two diagrams
shown in Figs.~\ref{diagram1}a and ~\ref{diagram1}b, which provide comparable contributions. 
However, as shown in Ref.~\onlinecite{prb07Glazman}, in graphene the Fock diagram dominates
over the Hartree diagram in the absence of magnetic field. This is a consequence
of the suppressed backscattering. {  We show\cite{appendix} that the weak magnetic field does not change the picture,  namely the Fock diagram is still dominating.}  
%In this paper we show that, in terms of sensitivity to
%a weak magnetic field, the Fock diagram also yields a dominant contribution.
We will thus focus on the sensitivity of the Fock diagram
to a weak magnetic field.

%%As was shown in Ref.~\onlinecite{prb07Glazman} the Fock diagrams dominate over Hartree diagrams because of the suppressed backscattering in graphene. We show that the same scenario survives in the weak magnetic field and evaluates these diagrams at finite $B$. We show that the phase $\varphi(r)\hat{\Sigma}_z$ gives rise to novel and unexpected transport and thermodynamic phenomena.

We start with a matrix generalization of the analytical expressions for
%Hartree and
the Fock diagram, Fig.~\ref{diagram1}a.
%expressing
%the interaction correction to the tunnel density of states.
For this purpose, we consider a non-magnetic impurity causing a perturbation
%given by
%described the potential
$\hat{u} \delta(\mathbf{r})$ and the screened interaction potential,
$U(\mathbf{r})$, with a radius $\sim k_F^{-1}$. The corresponding
expression reads

%analytical expressions of diagrams shown in Figs.~\ref{diagram1}a and ~\ref{diagram1}b, for tunneling DOS discussed in Ref.~\onlinecite{prb07Glazman}. We consider a non-magnetic impurity, given by the potential $\hat{u} \delta^2(\mathbf{r})$ and the screened interaction potential, $U(\mathbf{r})$. {\color{red}  The screening length has the characteristic scale, $\sim k^{-1}_F$}. The combined expression representing both Hartree and Fock contributions to the causal Green function is given by

%\begin{eqnarray}
% \label{HF}
% \delta G_\omega(\mathbf{r},\mathbf{r})=&&  \int d{\bf r}_1 d {\bf r}_2    G_{\omega}( \mathbf{r},\mathbf{r}_1)H_{\text{HF}}(\mathbf{r}_1,\mathbf{r}_2)G_{\omega}(\mathbf{r}_2,\mathbf{0} )\nonumber\\
% &&\times\hat{u} G_{\omega}(\mathbf{0} ,\mathbf{r})+\left(\hat{u}\leftrightarrow H_{\text{HF}}\right).
%  \end{eqnarray}

\begin{eqnarray}
\label{HF}
\delta G_\omega(\mathbf{r},\mathbf{r})=&&  \int d{\bf r}_1 d {\bf r}_2    G_{\omega}( \mathbf{r},\mathbf{r}_1)H_{\text{F}}(\mathbf{r}_1,\mathbf{r}_2)G_{\omega}(\mathbf{r}_2,\mathbf{0} )\nonumber\\
&&\times\hat{u} G_{\omega}(\mathbf{0} ,\mathbf{r})+\left(\hat{u}\leftrightarrow H_{\text{F}}\right).
\end{eqnarray}
Here $G_\omega$ is the free Feynman propagator of the Dirac electrons between the position
of impurity ${\bf r}=0$ and the point {\bf r}),
% %$H_{\Text{Hf}}=H_{\Text{H}}+H_{\Text{F}}$ With $H_{\Text{H}}=-I\Int (2\Pi)^{-1}  D\Omega
% \Text{Tr} [\Hat{U}G_\Omega(\Mathbf{R_1},\Mathbf{0}  )G_\Omega(\Mathbf{0} ,\Mathbf{R_1} )] U(\Mathbf{R_1}) \Delta(\Mathbf{R_1}-\Mathbf{R_2})
% $
while $H_F$ stands for nonlocal Fock potential
\begin{align}
\label{Fock}
H_F=\frac{i}{2 \pi}\int d\Omega G_{\omega+\Omega}(\mathbf{r}_1,\mathbf{0})
\hat{u}G_{\omega+\Omega}(0, \mathbf{r}_2) U(\mathbf{r_1}-\mathbf{r}_2).
\end{align}
%representing the Hartree and Fock contributions respectively.
The interaction correction to the local density of states, $\delta\nu_\omega({\bf r})$,
is related to the retarded Green's function
as $\delta\nu_\omega({\bf r})=-\frac{2}{\pi}\text{Tr~Im}~\delta G_{\omega}({\bf r},{\bf r}). $

%\cite{appendix}. {\color{red}
The structure of Eqs. (\ref{HF}), (\ref{Fock}) suggests that $\delta\nu_\omega({\bf r})$ contains
the product of  $4\times4$ matrices.
%Thus, one finds that corrections to the DOS can be reduced to products of $4\times4$ matrices.
%The Hartree diagram gives a sub-leading correction to DOS compared to the Fock diagram, due to the suppression of back-scattering in graphene. Thus we only focus on the evaluation of Fock diagram.}
%Namely, the correction from Hartree involves\begin{eqnarray} && H\equiv\text{tr}[\hat{u}M(\mathbf{r},+) M(-\mathbf{r},+)]\times \text{tr}[\hat{u}M(\mathbf{r},-) M(-\mathbf{r},-)]\nonumber  \end{eqnarray} For a qualitative discussion, let us now take $\hat{u}$ as the identity matrix. Upon using Eq.~\ref{3}, one finds\cite{appendix} $\text{tr}[M(\mathbf{r},s) M(-\mathbf{r},s)]\propto i(k_Fr)^{-1}+\omega^2_0 v_F^{-2} r^2 $. Thus Hartree contribution can only carry the leading field-dependent contribution $\sim \omega^2_0 v_F^{-2} r.$
In the semiclassical limit, all trajectories ${\bf r}\rightarrow {\bf r}_1\rightarrow {\bf r}_2\rightarrow 0\rightarrow {\bf r}$ contributing to $\delta G$ are close
to a straight line. With screened Coulomb potential being point-like,
%contributing  while the screened Coulomb potential is point-like.
%{\bf Then, the contribution Eq. corresponds to a particle being injected at r, moving to the impurity located at
%the origin, being scattered to r1 where it experiences the
%nonlocal Fock potential, and finally returning from r2 to the
%injection point r.}
the Fock diagram involves the following product of the $M$-matrices
\begin{eqnarray}
F\equiv\text{tr}\left[\hat{u}M(\mathbf{r},+) M(-\mathbf{r},-)\hat{u}M(\mathbf{r},-) M(-\mathbf{r},+)\right].
\end{eqnarray}
For a qualitative discussion, let us choose $\hat{u}$ in the form of a scalar, $u_0 \hat{I}$.
%now take $\hat{u}$ as $u_0 \hat{I}$.
Then the leading field-dependent term emerges as a coefficient in front of the product of the projection operators
$\text{tr}\left[\hat\Sigma_z\hat\Sigma_{x/y}\hat\Sigma_z\hat\Sigma_{x/y}\right]$.
Since the term $\hat\Sigma_z$ appears in the matrix $M$ in combination with $\varphi(r)$, we have $F\propto \varphi^2(r)$. With the help of the commutation relations for ${\hat\Sigma_x}$, ${\hat\Sigma_y}$, and ${\hat\Sigma_z}$, it is easy
to check that  $\text{tr}\left[\hat\Sigma_z\hat\Sigma_{x/y}\hat\Sigma_z\hat\Sigma_{x/y}\right]= -\text{tr}\left[I\right]$, i.e. it is nonzero.
An estimate for $F$ is $\sim u_0^2\varphi^2(r)\sim u_0^2\omega_0^2r^2/v_F^2$. With
characteristic $r$ being $v_F/\omega$, this estimate translates into $u_0^2\omega_0^2/\omega^2$. Below we examine a number of observables
having the structure similar to Eq. (\ref{HF}).

{\em Emerging zero-bias anomaly.}
For the scalar impurity scattering,  $\hat{u}=u_0\hat{I}$, there
is no zero-bias anomaly in graphene.\cite{prb07Glazman}
To convert the above estimate for $F$ into the $B$-dependent
correction to the density of states, we perform the
spatial averaging of Eq. (\ref{HF}), which  generates the impurity concentration, $n_i$.
Final result reads

\begin{eqnarray}
\label{8}
\frac{\delta\nu_\omega(B)-\delta\nu_\omega(0)}{\nu_F}\simeq   \frac{ \lambda_0 n_i u_0^2}{8\pi v^2_F  }   \frac{\omega_0^2}{\omega^2},
\end{eqnarray}
where $\lambda_0= k_F U_0  /(2\pi v_F)$ stands for dimensionless
interaction parameter, $U_0$ is the interaction potential with zero momentum transfer and $\nu_F=k_F/(\pi v_F)$. 

%%In Ref.~\onlinecite{prb07Glazman}, it was shown that there is no zero-bias anomaly for the scalar impurity scattering, $\hat{u}=u_0\hat{I}$. Here we show that the magnetic field dramatically enhances impurity scatterings, and gives rise to an emerging zero-bias anomaly.

\begin{figure}
	\centering
	\includegraphics[scale=0.5]{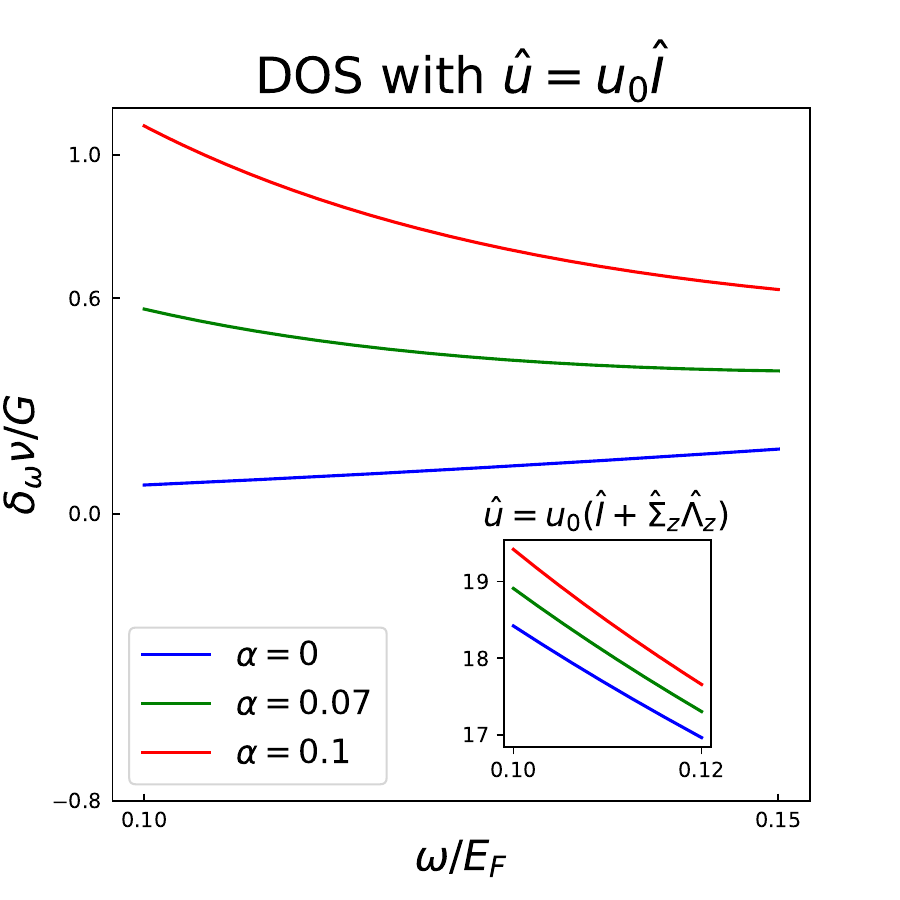}
	\caption{ (Color online)
		Plot (a) and the inset illustrate the energy dependence of the interaction correction to the density of states.
		Three curves correspond to the three values of the dimensionless magnetic field  $\alpha=(k_Fl)^{-2}$.
		Plot (a) is for the scalar impurity with magnitude $\hat{u}=u_0\hat{I}$.
		The correction, $\delta \nu$, is measured in the units of
		$G=\nu_F n_i ( u_0/2v_F)^2 \lambda_0/2\pi$.
		Note that for $\alpha=0$ the zero-bias anomaly is absent, so that $\delta \nu$ is a smooth function of energy, $\omega$, measured from the Fermi level.
		%(a)	 The first figure plots the interaction correction to the averaged density of states, $\delta_\omega \nu/G$, when the impurity potential is $\hat{u}=u_0\hat{I}$. Here $G=\nu_F n_i ( u_0/2v_F)^2 \sqrt{ \lambda/ 2\pi }$. In the absence of atomically sharp defects, the logarithmic anomaly vanishes and only smooth dependence of $\omega$, in fact a quadratic function $\sim \omega^2$, exists for $\alpha=0$.
		In the low-energy regime, $\omega/E_F<\sqrt{\alpha}$, the $B$-dependent anomalous term in
		$\delta\nu$ dominates and behaves as $\sim \alpha^2 E^2_F/\omega^2$.
		%DOS dominates, which behaves in an algebraic manner $\sim \alpha^2 E^2_F/\omega^2$.
		%Plot (b) is for the impurity-induced perturbation $u=u_0(\hat{I}+\hat\Sigma_x \hat \Lambda_x)$. For this perturbation, the logarithmic zero-bias anomaly is present even in the absence of
		%magnetic field.
		The inset of Plot (a) is for the impurity-induced perturbation
		%  The second figure plots DOS when $u-u_0\hat{I}\neq 0$. Here we take an example with $u=u_0(\hat{I}+\hat\Sigma_x \hat \Lambda_x)$.
		For this perturbation, zero-bias anomaly exists even in the absence of magnetic field.
		The magnetic contribution yields only a small correction to the logarithmic $\delta\nu$.
		% (c) The third plot In plot (c)  the temperature-dependent interaction correction to the effective velocity, $\delta_T v^*=v_T^*(B)-v_{T=0}^*(B)$ is shown. The inset shows the $B$-dependent component of the effective velocity, $\delta_T v^*(B)-\delta_T v^*(0)$. This part behaves as an inverse temperature, $\sim \alpha^2 E_F/T$.	
	}
	\label{DOS}
\end{figure}

%The Disorder Averaged Tunneling Dos In The Ballistic Regime Can Be Evaluated From The Imaginary Part Of Eqs.~\Ref{Hf}. According To Previous Analysis, We Already Observe That The Field-Dependent Correction Carries The Form Of $U_0^2(\Omega_0/\Omega)^2$. Here We Report The Complete Expression

The most general form of the point-like perturbation, $\hat{u}$, consistent with time-reversal symmetry is
%The time-reversal-symmetry-allowed matrix $\hat{u}$ in the potential can be generally %written as
$\hat{u}=u_0\hat{I}+\sum_{s,l=x,y,z} u_{sl } \Sigma_s \Lambda_l
$.  Here $\Lambda_{x,y}=\hat{\tau}_{x,y} \otimes \hat{\sigma}_z, \Lambda_z=\hat{\tau}_z \otimes\hat{\sigma}_0$.
%The $u_0\hat{I}$ is Coulomb-type impurity, representing the averaged static potential over the unit cell.
The remaining nine types of the disorder can be incorporated into Eq.~\ref{8} by
replacing $u^2_0$ by $t=u_0^2-\sum_{l}u^2_{zl}$.

%$t\equiv u_0^2-\sum_{l}u^2_{zl}$
%impurities are all atomically sharp impurities. With the general symmetry allowed matrix, %$u^2_0$ in Eq.~\ref{8} is replaced by $t\equiv u_0^2-\sum_{l}u^2_{zl}$.

The result Eq.~\ref{8} was obtained under the assumptions  $\omega \tau \gg 1$ and $\omega \gg \omega_0$
which ensure the ballistic regime and the irrelevance of the Landau quantization, respectively.

Emergence of a zero-bias anomaly in graphene in the presence of magnetic field manifests
itself in the local density of states (DOS), $ \delta\nu_\omega(\mathbf{r},B)= -2\pi^{-1}     \text{ tr}\left[\text{Im } G_R(\mathbf{r},\mathbf{r},\omega) \right]$.
Evaluation of Eq. (\ref{HF}) yields
\begin{eqnarray}
\label{9}
\frac{\delta\nu_\omega(\mathbf{r},B)-\delta\nu_\omega(\mathbf{r},0)}{\nu_F}\simeq \frac{ \lambda_0 t \omega_0^2   }{( 2\pi  v_F^2)^2 }   \cos \frac{\omega r}{v_F} .
\end{eqnarray}
Note that, unlike the $B=0$ case,\cite{prb07Glazman} the interaction correction Eq. (\ref{9}) is isotropic. The most dramatic difference between Eq. (\ref{9}) and the
$B=0$ result is that the zero-field correction falls off as $1/r^2$, while the amplitude of oscillations in Eq. (\ref{9}) does not depend on $r$. Naturally, the fall-off starts from the distances
$r  \gtrsim R_L =v_F/\omega_0$,
where Eq. (\ref{9}) does not apply. Technically, the extra factor $r^2$ comes from $\varphi^2(r)$ in the factor $F$.
In relation to the local DOS, we would like to point out that
%this local DOS
it can be measured experimentally using the scanning tunneling microscopy (STM)\cite{prb07Marchini,prl09Andrei}.

%Eq.~\ref{9} is obtained for $k^{-1}_F\ll r\ll R_L$.

%

{\em  Quasi-particle lifetime}. Energy dependence of electron-electron scattering rate, $\tau_{ee}^{-1}$,
in doped graphene is $\omega^2\ln\left(E_F/\omega\right)$, as in a regular Fermi liquid.\cite{prb07das} This dependence emerges already in the lowest order of the perturbation theory. Corresponding diagram is illustrated in Fig.~\ref{diagram1}. Subsequent summation of the higher-order diagrams within the random-phase approximation (RPA) modifies the prefactor in $\tau_{ee}^{-1}$. Equally, the calculations leading to non-analytic interaction corrections\cite{prb03Chubukov} apply to the doped graphene.
With regard to the magnetic field dependence of $\tau_{ee}^{-1}$, it appears that,  similarly to the zero-bias anomaly, the leading $B$-dependence originates from the Fock diagram on Fig.~\ref{diagram1}c, which is beyond the RPA.

%lifetime
%The doped graphene is a 2D Fermi-liquid\cite{prb07das}. As such,
%it may develop non-analytic interaction corrections\cite{prb03Chubukov} to observable characteristics of the system. Here, we concentrate on the analytical study of the field-dependent correction to the quasi-particle lifetime.
%
%The leading contribution to the quasi-particle lifetime comes from two second order in interactions Feynman diagrams, depicted in  Fig.~\ref{diagram1}. A well-known result in 2D electron gas is that each order of perturbation theory respect to the interaction contributes leading $\omega^2 E_F^{-1} \ln( E _F \omega^{-1})$ correction to the lifetime. Thus, the non-perturbative result for $\tau$ has been obtained upon the summation of Random Phase Approximation (RPA) series\cite{prb82Quinn,prb96Sarma}. We observe that in the presence of magnetic field, field-dependence of these RPA diagrams is sub-dominant. The main field-dependent contribution to the lifetime is generated by non-RPA diagrams, with the leading contribution being generated by the Fock diagram on Fig.~\ref{diagram1}c.

The result for the correction, $\delta \tau_{ee}^{-1}(B)$, depends on the
ratio $\omega/T$. In the low-$T$ limit, $\omega \gg T$, this correction reads
\begin{eqnarray}
\label{fock}
%\tau^{-1}(B)-\tau^{-1}(0)
\delta \tau_{ee}^{-1}(B) \simeq \frac{ \lambda_0 \lambda_{2k_F} \omega_0^2}{2\pi E_F}  \ln\Big(\frac{|\omega|}{\Delta}\Big),~~
%\quad \text{if}\quad
|\omega|\gg T,
\end{eqnarray}
where $\lambda_{2k_F}= k_F U_{2k_F}  /(2\pi v_F)$, $\Delta=\text{max}\{T,\tau^{-1}_{ee}\}$. The relative magnitude of the correction is essentially $\left(\omega_0/\omega\right)^2$
and, similarly to the zero-bias anomaly, it originates from
the magnetic phase $\hat{\Sigma}_z \varphi(r)$ of the propagator in Feynman diagrams.

%Here $\tau(B)$ is the quasi-particle lifetime in the presence of magnetic field $B$. The result indicates that the energy dependence of the field-dependent correction to the lifetime is {\em purely logarithmic}. We note here that at a finite $B$, the $\omega$-dependence in lifetime is significantly enhanced and more singular. The result is obtained employing the magnetic phase $\hat{\Sigma}_z \varphi(r)$ of the propagator in Feynman diagrams. Qualitatively, the field-dependence of lifetime comes from the small factor $\sim\omega_0^2/\omega^2$ in front of RPA lifetime. %The origin of the small factor is the following. Since $\textbf{tr} \hat{\Sigma}_z=0$ and $\textbf{tr} \hat{\Sigma}^2_z\neq0$, the leading term in the expansion of $\exp\left( -i\text{sgn}(\omega)\varphi(r)\hat{\Sigma}_z\right)$ is  $\varphi^2(r)=r^2 (2k_F l^2)^{-2}$. The characteristic length of $r$ is the scale $\sim v_F/\omega$. This in turn yields the small factor, $\omega^2_0/\omega^2 $.

In the high-temperature limit, $T\gg\omega$, evaluation of the
$B$-dependent correction to the diagram Fig.~\ref{diagram1}c yields
\begin{eqnarray}
\label{fock1}
%\tau^{-1}(B)-\tau^{-1}(0)
\delta \tau_{ee}^{-1}(B)\simeq -  \ln (2) \frac{ \lambda_0 \lambda_{2k_F} \omega_0^2}{2\pi E_F},~~ T \gg |\omega|.
\end{eqnarray}
Note that the correction is $T$-independent, but  it exists on the background of the $T^2$ main term.

%Namely, the logarithmic behavior disappears. Note that the obtained correction is smaller than the $\tau^{-1}(0)$ and is {\em independent} of $T$.

%This result is beyond expectations. Usually, magnetic effects are expected vanish in an exponential manner, $\exp [-2\pi T/\omega_0]$, at the limit $T/\omega_0\gg 1$. Qualitatively, the phase $\varphi(r) \Sigma_z$ changes the integrand to be $T^2 \varphi^2(r) (E_F r)^{-1} \cosh^{-2}(\pi T r /v_F)$. Still, the integral mainly comes from the region where $r<v_F/T$. In the region, the integrand is approximated by $(T/v_F)^2 \omega^2_0 r/E_F$.Then, the integral of $(T/v_F)^2 \omega^2_0 r/E_F$ on the $(0,v_F/T)$ erases the dependence of $T$ and yields $\omega_0^2/E_F$, a constant correction in Eq.~\ref{fock1}.

\begin{figure}
	\centering
	\includegraphics[scale=0.55]{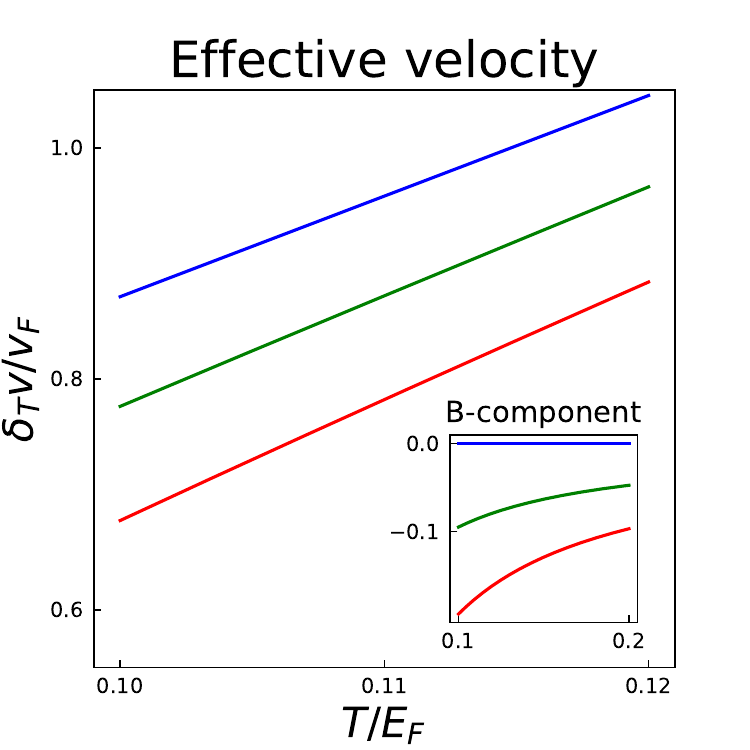}
	\caption{ (Color online)
		In plot
		the temperature-dependent interaction correction to the effective velocity, $\delta_T v^*=v_T^*(B)-v_{T=0}^*(B)$ is shown. The inset shows the $B$-dependent component of the effective velocity, $\delta_T v^*(B)-\delta_T v^*(0)$. This part behaves as an
		inverse temperature, $\sim \alpha^2 E_F/T$.	
	}
	\label{velocity}
\end{figure}
{\em Effective velocity and specific heat.}
In the doped graphene, as in 2D electron gas, the effective velocity of quasi-particles, $v^*$, and specific heat, $C_v$, are expected to acquire interaction corrections\cite{prb03Chubukov, prb07das,appendix}.
These corrections scale as $\delta v^* \propto T$ and $\delta C_v\propto T^2$, respectively. Both anomalies originate from the non-analytic corrections to the quasi-particle lifetime \cite{prb03Chubukov}.
Here we trace how the $\omega_0^2$-corrections specific for graphene manifest themselves
in $v^*$ and $C_v$. The question of interest is the temperature dependence of these
corrections. We found that the correction to $v^*$ behaves as $\omega_0^2/T$, while the
correction to $C_v$ is $\propto \omega_0^2/v_F^2$ and is $T$-independent. Both originate from $\omega_0^2$ correction to the lifetime given by Eqs.~\ref{fock} and ~\ref{fock1}.

%$v^*$ acquires a correction scaling inversely with temperature, $\sim \omega_0^2 (k_F T)^{-1}$ and $C_v$ obtains a temperature-independent contribution, $\sim \omega_0^2v_F^{-1}$. These two corrections originate from magnetic components in lifetimes, given by Eqs.~\ref{fock} and ~\ref{fock1}.
%The interaction correction to the spectrum of quasi-particles comes from the real part of the self energy. Here we consider the diagrams, shown in Figs.~\ref{diagram1}c and ~\ref{diagram1}d. One can recover\cite{appendix} that
%the Hartree diagram in Fig.~\ref{diagram1} gives the leading field-independent correction to the spectrum: $\delta \epsilon\propto \omega |\omega|/E_F $, if $\omega \gg T$ and $\delta \varepsilon\propto \omega T/E_F $, if $\omega \ll T$. Here $\delta \varepsilon$ is the interaction correction to the spectrum. While the magnetic field does not modify the Hartree correction,

Another ingredient required to find the $B$-dependent corrections to $v^*$ and $C_v$
is the electron spectrum renormalized by the interactions. The corresponding
$\omega_0^2$-correction comes from the Fock diagram Fig.~\ref{diagram1}c
%we find that the Fock diagram Fig.~\ref{diagram1}c gives the leading magnetic corrections
\begin{eqnarray}
\label{4}
\text{Re}[\Sigma(\omega,B)- \Sigma(\omega,0)] \simeq -  \frac{ \lambda_0 \lambda_{2k_F} \omega_0^2}{16E_F} \times
\left\{
\begin{array}{cc}
\text{sgn}(\omega) , &\text{$|\omega| \gg T$}\\
\omega/(2T) ,&\text{$|\omega| \ll T$}.\\
\end{array}
\right. \nonumber\\
\end{eqnarray}
%Here the $\delta\varepsilon(B)$ is the interaction correction to the spectrum of quasi-particles under the magnetic field $B$.
The above correction can, in principle,
be measured using the Angle-resolved photoemission spectroscopy (ARPES)\cite{Lv2019}
from the analysis of  the constant energy maps\cite{prb08M} at different values of $B$.

%We propose an observation of this effect using Angle-resolved photoemission spectroscopy (ARPES)\cite{Lv2019} by measuring constant energy maps\cite{prb08M} at different values of $B$ in the weak field limit.

%In graphene, electrons are massless and the velocity of quasi-particles is renormalized by interactions.

In the limit $T\gg \omega$, the renormalized spectrum  Eq.~\ref{4} %renormalizes the
leads to the following correction to the effective velocity of {  quasi-particles $v=v_F/\left(1-\partial_\omega \text{Re}\Sigma|_{\omega=0}\right)$,\cite{appendix}}
\begin{eqnarray}
\frac{v(B)-v(0)}{v_F}\simeq      -  \frac{ \lambda_0 \lambda_{2k_F} \omega_0^2}{32 E^2_F}\frac{E_F}{T}.
\end{eqnarray}
%Here $v_T(B)$ is the effective velocity of quasi-particles at temperature $T$ and field $B$.
Note that $v(0)$ contains a non-magnetic interaction correction which is linear in $T$. On the contrary, the $B$-dependent correction is $\propto T^{-1}$. This feature is illustrated in Fig.~\ref{velocity} for several values of $B$.
Since the thermodynamical potential, $\Omega$, involves the summation over energies of quasi-particles near the Fermi level, the energy correction in Eq.~\ref{4} has non-trivial implications for thermodynamics. Here we consider the specific heat per unit volume, $C_v=V^{-1} \partial \Omega/\partial T$, where $V$ is the volume of system. The result for specific heat\cite{appendix} in the limit $T\gg \omega_0$
is the following,
\begin{eqnarray}
\label{6}
\delta C_v(B)-\delta C_v(0)\simeq -  \frac{\lambda_0 \lambda_{2k_F}\omega_0^2}{ 8\pi v_F^2},
\end{eqnarray}
where $\delta C_v(B)$ is the interaction correction to the specific heat.
Note that $\delta C_v(0)$ contains the conventional $T^2$ term, specific for 2D Fermi liquid. We find that the field-dependent correction to $\delta C_v$ is a $T$-independent.  In the absence of electron-phonon interactions, the field-dependent correction exists in a parametrically large interval of temperatures, $\omega_0<T<E_F$. Eq.~\ref{6} can be verified experimentally by measuring the specific heat of graphene in a comprehensive Raman optothermal method\cite{17nanoli}.

{\em Conclusion.} Our main finding is that, for two-dimensional Dirac electrons, application of a
weak magnetic field enhances significantly the many-body effects. This is unlike the conventional
2D electron gas.
%We showed that the many-body properties of the Dirac electrons in a weak magnetic field are significantly enhanced compared to those in a 2D electron gas.
The reason for this is the pseudospin-dependent magnetic correction in Dirac electron propagators, $\sim \hat\Sigma_z\varphi(r)$. For many-body effects to unfold, the energy, $\omega$, measured from the Fermi level
should exceed $\omega_0= v_F(k_F l^2)^{-1}$, which is the inter-Landau-level distance at the Fermi level.
We have only considered the low-temperature properties of interacting electrons in the doped graphene, so
that the interaction with phonons\cite{prb21AS,prb10Faugeras} can be neglected.

%%thermodynamics of the electrons.
%Interaction with phonons may give rise
%to a new energy scale for temperature, enriching thermodynamic characteristics.

Our predictions for observables given by Eqs.~\ref{fock}-\ref{6} and by Eq. \ref{8}
all emerge as a result of evaluation of the Fock diagrams illustrated in Figs. \ref{diagram1}a, \ref{diagram1}c.
It is nontrivial that, while these diagrams are not leading and even do not belong to the RPA sequence,
they are responsible for the sensitivity to a weak magnetic field. Importantly, the higher-order diagrams,while leading to the renormalization of the interaction vertex, do not modify thepredicted  $\omega,T$-dependencies.

%We have estimated the effect of loop-renormalization of the interaction vertex in Figs.\ref{diagram1}a, \ref{diagram1}c and verified that the leading field-depend effects, and their $\omega,T$-dependencies in Eqs.~\ref{fock} through \ref{6} and Eqs. \ref{8} reported in this work are not being modified. This implies that the reported interaction effects have a non-perturbative origin and are beyond RPA.

Other origin of the
$B$-dependence of the interaction effects is either spin via the Zeeman splitting
coming from spin or
%of the spin origin, i.e. Zeeman splitting,
%or of the orbital origin, which
or the orbital effect via the curving of the electron trajectories in magnetic field.
We have checked\cite{appendix} that these two mechanisms lead to the $B$-dependent
corrections which are sub-leading compared to the ones originating from the pseudospin-dependent phase of Dirac propagators.

{  Finally, we emphasize that our results apply for the doped graphene, where the Fermi energy is far away from the neutrality. The condition $\omega\gg \omega_0$ in the present letter is automatically violated at neutrality. The question about $\nu=0$ Landau level is interesting and remains open\cite{prb12Maxim, rmp11Goerbig}. 
	
}
%induce field-dependent interaction corrections to the thermodynamics and the DOS that are sub-leading\cite{appendix} compared to the ones originating from the pseudospin-dependent orbital phase of Dirac propagators
%
%We also consider field-dependent interaction effects originating from the Zeeman effect and curving of trajectories under B-field. We find that these two mechanisms induce field-dependent interaction corrections to the thermodynamics and the DOS that are sub-leading\cite{appendix} compared to the ones originating from the pseudospin-dependent orbital phase of Dirac propagators

{\em Acknowledgements}. The research was supported by startup funds from the University of Massachusetts, Amherst (K.W. and T.A.S.), and by the Department of Energy, Office of Basic Energy Sciences, Grant No. DE-FG02-06ER46313 (M.E.R.).

 	\bibliography{document}

\begin{thebibliography}{51}
\bibitem [{foo()}]{footnote1}%
  \BibitemOpen
  \href@noop {} {\bibinfo  {journal} {To visualize the chiral symmetry breaking
  around a single valley, we consider the (pseudo)spin vector,
  $\mathbf{v}_{Q}(\mathbf{r})$, of the wavefunction at $Q=K$ or $K'$, defined
  as follows: we take the positive eigenstate $\psi(\mathbf{r})$ of the matrix
  $M$ in Eq.~\ref{3}, and project $\psi(\mathbf{r})$ into $K$-valley (or
  $K'$-valley) as $\psi_{K}(\mathbf{r})$ and then define
  $\mathbf{v}_{K}(\mathbf{r})=\langle \psi_{K}(\mathbf{r}) |
  (\hat\sigma_x,\hat\sigma_y,\hat\sigma_z) |\psi_{K}(\mathbf{r}) \rangle$.
  Since the chiral transformation connects the two valleys, as shown in
  Fig.~\ref{pictorial}), the spin vectors at $K$, $K'$ points satisfy the
  property $\mathbf{v}_{K'}(\mathbf{r})=\mathbf{v}_{K}(-\mathbf{r})$. In
  Fig.~\ref{pictorial}, we plotted the vector $\mathbf{v}_K(\mathbf{r})$ to
  illustrate the effect of chiral-symmetry breaking around $Q=K$}\
  }\BibitemShut {NoStop}%
\end{thebibliography}
 	
 	\pagebreak
 	\widetext
 	\begin{center}
 		\textbf{\large Supplemental Material: Interaction effects in Graphene in a weak magnetic field }
 	\end{center}
 	%%%%%%%%%% Merge with supplemental materials %%%%%%%%%%
 	%%%%%%%%%% Prefix a "S" to all equations, figures, tables and reset the counter %%%%%%%%%%
 	\setcounter{equation}{0}
 	\setcounter{figure}{0}
 	\setcounter{table}{0}
 	\setcounter{page}{1}
 	\makeatletter
 	\renewcommand{\theequation}{S\arabic{equation}}
 	\renewcommand{\thefigure}{S\arabic{figure}}
 	\renewcommand{\bibnumfmt}[1]{[S#1]}
 	\renewcommand{\citenumfont}[1]{S#1}

\section{Dirac propagator in the presence of a weak magnetic field}
This section provides a derivation of the Dirac propagator for the $K$-valley using the operator formalism.
%(summations over Landau levels).
In the presence of magnetic field, the spectrum of the
Dirac electrons transforms into the ladders of Landau levels.
%The positive Landau energy levels are given by
The level positions are given by $\omega_n=\sqrt{2n} v_F/l$, where $v_F$ is the Fermi velocity and $l$ is the magnetic length.
The corresponding wavefunctions in the $K$-valley are given by $\psi_{n,k_x}(\mathbf{z})=\frac{1}{\sqrt{2}}\Big(\varphi_{n-1,k_x}(\mathbf{z}),- \varphi_{n,k_x}(\mathbf{z})\Big)$, where $\varphi_{n,k_x}(\mathbf{z})$ is the wavefunction of $n^{\text{th}}$ Landau level of the  2D electrons.

%Considering the positive Fermi energy $E_F$, we only need to consider the Landau levels around
To calculate the propagator, it is sufficient to consider the group of levels around $E_F$.
Then the definition of the propagator leads to the following starting expression:
\begin{eqnarray}
G_K^{s,s'}(\mathbf{z},\mathbf{z'};\omega)\simeq\int \frac{dk_x}{2\pi} \sum_{n} \psi^s_{n,k_x} (\mathbf{z}) \psi^{s',*}_{n,k_x}(\mathbf{z'})
\times \frac{1}{\omega- \omega_n+i\delta \Theta(\omega-E_F)}.
\end{eqnarray}
Here $\mathbf{z}=(x,y)$, $\mathbf{z}'=(x',y')$, while the indices $s$ and $s'$ taking the values $s,s'=\pm $
refer to the $A/B$ sublattices.

The off-diagonal propagators could be expressed in terms of diagonal ones via the following expression
$
G^{s,s'}_K(\mathbf{z},\mathbf{z'},\omega)=\frac{ l^2 \omega {p}_{s'}}{ v_F r^2 } \left[ G^{11}_K(\mathbf{z},\mathbf{z'},\omega)-G^{22}_K(\mathbf{z},\mathbf{z'},\omega)\right]  \label{16}
$
with $\tilde{p}_\pm= {\pm(y-y')-i(x-x')} $. Thus, we focus on the summation over the Landau levels for diagonal propagators. Upon substituting the expression of $\psi_{n,k_x}(\mathbf{x})$ and variable change $n\rightarrow n- (s+1)/2$, we obtain
\begin{eqnarray}
G_K^{s,s }(\mathbf{x},\mathbf{x'};\omega)\simeq \frac{1}{2}\int \frac{dk_x}{2\pi} \sum_{n} \varphi_{n,k_x} (\mathbf{x}) \varphi^{ *}_{n,k_x}(\mathbf{x'})
\frac{1}{\omega- \omega_{ n+ (s+1)/2}}.
\end{eqnarray}
Next we expand $\omega_{ n+ (s+1)/2}=\sqrt{2n+(s+1)}v_F/l$ around the Fermi energy and find
\begin{eqnarray}
\omega_{ n+ (s+1)/2} \simeq E_F \Big(1+\frac{s}{2k_F^2 l^2}\Big)+(\delta n+1/2 )\frac{ E_F }{ k_F^2l^2}.  \nonumber
\end{eqnarray}
Here we introduce an effective Fermi energy $E_F (1+s(2k_F^2 l^2)^{-1})$ and effective cyclotron frequency $\omega_0=E_F/ k_F^2l^2$. Subsequently, the effective Fermi energy introduces an effective momentum as $k_F^{s,s}=E^s_F/v_F$. One readily finds:  $k_{s,s}=k_F^{s,s} (1+s(2k_F^2 l^2)^{-1})$.

The effective momentum in the off-diagonal Green functions is exactly the Fermi momentum. Thus, we can keep the off-diagonal propagators the same as the ones {\it without} magnetic field. The diagonal Dirac propagators can be identified with the 2D electron gas propagator with the effective momentum $k_F^{s,s}$. Upon the use of the expression of the propagator of the 2D electron gas\cite{prl07Sedrakyan}, one finds
\begin{eqnarray}
\label{chi}
G(\mathbf{z},\mathbf{z}';\omega)\simeq e^{-i\chi } I(\mathbf{z}-\mathbf{z}';\omega)
\hat{M}(\mathbf{z}-\mathbf{z}';\text{sgn}(\omega)),
\end{eqnarray}
where $
I(\mathbf{z}-\mathbf{z}';\omega)= \frac{ 1 }{ 2v_F} \sqrt{\frac{k_F}{ 2\pi r }} \exp i\text{sgn}(\omega)\left(k_Fr+\frac{\omega}{v_F}r+ \frac{\pi}{4}-\frac{ r^3}{24k_F l^4} \right) . \nonumber $
Here $r=|\mathbf{z}-\mathbf{z'}|$, $\chi=(x-x')(y+y')/(2l^2)$ describe the breaking of the translational invariance. The matrix  reads as
\begin{eqnarray}
\label{S4}
{M}\left(\mathbf{r};\text{sgn}(\omega)\right)&\simeq&\Big(
\text{sgn}(\omega)+\frac{i}{2k_F r}
\Big) \hat{r}\cdot \boldsymbol\Sigma + \exp\Big\{-i\text{sgn}(\omega)\varphi(r)\hat{\Sigma}_z\Big\},
\end{eqnarray}
where $\hat{r}=\mathbf{r}/r$, $\Sigma_{x,y}=\sigma_{x,y}\otimes \tau_z$, $\varphi(r)=r/(2k_Fl^2)$ and $\Sigma_{z}=\sigma_{z}\otimes \tau_0$. Here $\sigma$ acts on the pseudo-spin space and $\tau$ acts on the valleys. Expanding the exponents
%$\exp$-function
up to $\varphi^2$, one recovers the expression of  Eq.~(2) from the main text. The result in Eq.~\ref{S4} applies when $k^{-1}_F<r<k_Fl^2$.

\section{Hartree v.s. Fock diagrams}
In the main-text, we argue that Fock diagram (which is a non-RPA diagram) gives leading field-dependent corrections while Hartree diagram contribution is subleading. This property comes from the fact that the back-scattering, which is the relevant process for
the Hartree diagram, is suppressed in the graphene. Here we provide a detailed calculation when one consider quasi-particle lifetime. For other physical quantities (e.g. )

\begin{itemize}
	\item When one use Hartree diagram to evaluate the quasi-particle life-
	time, one needs to consider the product of two dynamical polarization operator. The species with $2k_F$ momentum transfer in the Hartree diagram contains the following product of matrices
	\begin{eqnarray}
	\label{S5}
	H\equiv\text{tr}\left[M(\mathbf{r},+) M(-\mathbf{r},+) \right] \times \text{tr}\left[M(\mathbf{r},-) M(-\mathbf{r},-)\right].
	\end{eqnarray}
	Using $M$ in Eq.~\ref{S4}, one could recover the following equation
	\begin{eqnarray}
	\label{S6}
	\text{tr}\left[M(\mathbf{r},s) M(-\mathbf{r},s) \right]= 4\Big(
	-1-i\frac{s}{k_Fr}+ \cos \varphi(r)  
	\Big)\simeq -4\Big(
	i\frac{s}{k_Fr} +\frac{\varphi (r)^2}{2}
	\Big)
	\end{eqnarray}
	Note that the cancellation of leading term in  $- 1+\cos \varphi  $ is the result of the suppression of the back-scattering (and then the faster decaying term $1/(k_Fr)$ plays a important role in graphene). One may put Eq.~\ref{S6} into Eq.~\ref{S5} and obtain
	\begin{eqnarray}
	\label{S7}
	H\simeq 16  \Big(
	\frac{i}{k_Fr} +\frac{\varphi (r)^2}{2}
	\Big) \Big(
	-\frac{i}{k_Fr} +\frac{\varphi (r)^2}{2}
	\Big)
	\end{eqnarray}
	One can clearly observe that the crossing term, which is proportional to $\varphi^2 (\propto B^2)$, vanishes. Even if the crossing term does not vanish, the extra decaying power $r^{-1}$ in Eq.~\ref{S6} makes the cross term in Hartree's contribution sub-leading compared to Fock's one, which will be calculated below.
	\item In the Fock diagram, the forward scattering becomes the relevant
	process. The following product of matrices is involved
	\begin{eqnarray}
	F\equiv\text{tr}\left[M(\mathbf{r},+) M(-\mathbf{r},-)M(\mathbf{r},-) M(-\mathbf{r},+)\right].
	\end{eqnarray}
	Using $M$ in Eq.~\ref{S4}, one could recover the following equation
	\begin{eqnarray}
	M(\mathbf{r},+) M(-\mathbf{r},-)  =  2\Big(1+\hat{r}\cdot \boldsymbol\Sigma \exp \{i\varphi(r) \Sigma_z\}\Big)
	\end{eqnarray}
	Then put the equation above into $F$ and one obtains
	\begin{eqnarray}
	\label{s10}
	F=16\Big(1-\cos \varphi(r)\Big)\simeq 8 \varphi(r)^2
	\end{eqnarray}
	Thus one could observe that $F$ in Eq.~\ref{s10} give stronger contribution than $H$ in Eq.~\ref{S7}.
\end{itemize}

\section{Derivation of the DOS}
We start from the standard expression for the density of states, $\nu(\omega;B)$,
in terms of the retarded Green function
%relation of tunneling DOS and the retarded Green function:
\begin{eqnarray}
\nu(\omega;B)=-\frac{2}{\pi} \int d^2r \textbf{Im} \textbf{tr} G_R(\mathbf{r},\mathbf{r};\omega).
\end{eqnarray}
%Here $\nu(\omega;B)$ is the DOS and $\nu(\omega;B)$ is evaluated from single-particle Green function.
In the ballistic regime, the interaction correction to $\nu(\omega;B)$ is given by the Fock and the Hartree
diagrams depicted in Figs.~2a and 2c, respectively. Specifics of graphene is the absence of the backscattering.
As a result, the Hartree diagram, which is dominated by the backscattering, does not lead to the zero-bias anomaly.
% which manifests itself
%the
%
%The latter is calculated using the standard Feynman diagram approach. Motivated by the the search of the zero-bias anomaly in graphene, we consider the Fock diagram depicted in Fig.~2a of the main text
Analytical expression for the Fock diagram reads
\begin{eqnarray}
\label{fock_diagram}
&&\delta_f G(\mathbf{r},\mathbf{r};\omega)\simeq i2 U\int d^2r_1 \int_{-E_F}^{-\omega} \frac{d\Omega_1}{2\pi} G(\mathbf{r},\mathbf{r}_1;\omega)G(\mathbf{r}_1,0;\omega+\Omega_1)\nonumber\\
&&\times\hat{u}G(0, \mathbf{r}_1;\omega+\Omega_1)G( \mathbf{r}_1,0;\omega )\hat{u}G( 0,\mathbf{r};\omega ).
\end{eqnarray}
Here we consider the touching potential. Thus the interaction potential in momentum space is uniform, namely, a single number $U$.
%the zero momentum component of the interaction potential.
%\begin{eqnarray}\label{fock_diagram}&&\delta_fG(\mathbf{r},\mathbf{r};\omega)=i\int d^2r_1 \int \frac{d\Omega_1}{2\pi} G(\mathbf{r},\mathbf{r}_1;\omega)G(\mathbf{r}_1,0;\omega+\Omega_1)\nonumber\\&&\times \int d^2 r_2 \hat{u}G(0, \mathbf{r}_2;\omega+\Omega_1)G( \mathbf{r}_2,0;\omega )\hat{u}G( 0,\mathbf{r};\omega ){V}(\mathbf{r_1}-\mathbf{r_2}). \nonumber\end{eqnarray}
In Eq. (\ref {fock_diagram}) we assume that $\omega$ is positive and use the  Green function, $G$, instead of $G_R$.
%Here the function $G$ above is the free and causal Green function for Dirac electrons. When $\omega >0$, $\delta_f G(\mathbf{r},\mathbf{r};\omega)$ coincides with the corresponding correction to the retarded Green function.

Spatial averaging of $\delta_f G(r,r)$ is accomplished with the help of the following identity
\begin{eqnarray}
\label{r1}
\int d^2 r G(0,\mathbf{r};\omega)G(\mathbf{r},\mathbf{r}_1;\omega)=-\partial_\omega G(0,\mathbf{r}_1;\omega).
\end{eqnarray}
%Since we consider the semi-classical region, we could make use of the asymptotic expression and obtain
Another simplification comes from the fact that the distances contributing to the integral  Eq. (\ref {fock_diagram})
are large, so that the Green function can be replaced by the semiclassical asymptote

\begin{eqnarray}
\label{r2}
\partial_\omega G(0,\mathbf{r}_1;\omega)\simeq i\text{sgn}(\omega) \frac{r_1}{v_F} G(0,\mathbf{r}_1;\omega).
\end{eqnarray}
Subsequent steps are in line with the calculation in
%Ref. \onlinecite{prb07Glazman}.
Phys. Rev. B {\bf 76}, 165402 (2007).
They involve
substituting the asymptotic expressions for the Green functions into $\delta_f\nu(\omega;B)$,
calculating the product of matrices entering the Green functions and integrating out the
intermediate frequency $\Omega_1$. As a result, the expression for $\delta_f\nu(\omega;B)$
simplifies to
\begin{eqnarray}
\label{18}
\delta_f\nu(\omega;B)&=& -2 n_i U_0\textbf{Re} \int d\theta dr_1 \frac{k^2_F}{32\pi^4v^4_F} \frac{1}{ r_1} e^{2i\omega r_1/v_F}\times \textbf{tr}(\hat{u}^2-\hat{\Sigma}(\psi)\hat{u} \hat{\Sigma}(\bar{\psi}) \hat{u} ),
\end{eqnarray}
%Let us now confine the attention to $\omega>0$.
%After putting the asymptotic expression of the Green functions, one can calculate the products of matrices from products of two Green functions and integrate out frequency $\Omega_1$. Then one obtains the following integral expression for $\delta_f\nu(\omega;B)$
%\begin{eqnarray}
%\label{18}
%\delta_f\nu(\omega;B)&=& - U_0\textbf{Re} \int d\theta dr_1 \frac{k^2_F}{32\pi^4v^4_F v} \frac{1}{ r_1} e^{2i\omega r_1/v_F}\times \textbf{tr}(\hat{u}^2-\hat{\Sigma}(\psi)\hat{u} \hat{\Sigma}(\bar{\psi}) \hat{u} ).
%\end{eqnarray}
where $n_i$ is the impurity concentration.
In Eq. (\ref{18}) the angle $\theta$ is the angular coordinate of $\mathbf{r}_1$, while the angles $\psi, \bar{\psi} $
are defined as
\begin{equation}
\psi=\theta+ \varphi(r_1),\quad \bar{\psi}=\theta - \varphi(r_1).
\end{equation}
Finally, the function $\hat{\Sigma}$ is expressed via $\psi$, $\bar{\psi}$
as follows
\begin{equation}
\hat{\Sigma}(\psi)=(\cos\psi, \sin \psi)\cdot (\hat{\Sigma}_x, \hat{\Sigma}_y).
\end{equation}
Eq. (\ref{18}) illustrates how the magnetic phase, $\varphi(r_1)$, from the electron
propagator enters the interaction correction to the density of states.

%$\hat{\Sigma}(\psi)$ are defined by
%\begin{eqnarray}
%\label{19}
%\psi=\theta+ \varphi(r_1),\quad \bar{\psi}=\theta - \varphi(r_1),\quad
%\hat{\Sigma}(\psi)=(\cos\psi, \sin \psi)\cdot (\hat{\Sigma}_x, \hat{\Sigma}_y)
%\end{eqnarray}
%Now we see how the magnetic phase, $\varphi(r_1)$, from the electron propagator enters into the expression for
% the averaged DOS.

To explore the magnetic field dependence, we analyze the intermediate integral
\begin{eqnarray}
\int \frac{d\theta}{2\pi} \textbf{tr}( \hat{u}^2-\Sigma(\psi) \hat{u} \Sigma(\bar{\psi}) \hat{u} )\simeq 4\sum_{l=x,y,z} (2 u^2_{zl}+ u^2_{yl}+ u^2_{xl}) +t \frac{ 2r_1^2}{ k_F^4l^4}.
\end{eqnarray}
Here $t= u^2-\sum_{l}u^2_{zl} $. From the  above expression, one concludes that only the scalar potentials, including $u, u_{zz}, u_{zx}$ and $u_{zy}$, are sensitive to the magnetic field. Defining a dimensionless variable $x=2 \omega r_1/v_F$, we cast the magnetic-field correction in the form of a single integral over $x$
\begin{eqnarray}
&&\label{21} \frac{\delta_f\nu(\omega;B)-\delta_f\nu(\omega;0)}{\nu_F} = - \frac{k_F U_0}{v_F} \frac{ t}{16\pi^2 v^2_F v} \frac{ \omega_0^2}{ \omega^2} \int_{2\omega/E_F}^{2\omega /\omega_0} x e^{-\epsilon x} \cos x dx.
\end{eqnarray}
Here $\epsilon$ is introduced as a cutoff. In the limit $\omega_0\ll \omega \ll E_F$, we get
%Here $\epsilon= \delta/2\omega\ll 1$. Using $\delta=1/2\tau$, where $\tau$ is the lifetime time of quasi-particles,  at $\omega/E_F\ll 1$ and the weak field limit $\delta/\omega_0\gg 1$, we obtain
\begin{eqnarray}
\label{anomalye}
\frac{\delta_f\nu(\omega;B)-\delta_f\nu(\omega;0)}{\nu_F}\simeq n_i\frac{k_F U_0}{v_F}\frac{t}{16\pi^2 v^2_F} \frac{ \omega_0^2}{ \omega^2} \label{42} .
\end{eqnarray}
We thus arrive to Eq. (8) of the main text.
%The summation over all impurities in the sample transforms the inverse of volume $1/v$ into $n_i$, the density of impurities. This is the main conclusion about DOS in the main text.

\section{Self-energy calculation at finite temperature}
In this section, we provide a detailed calculation of the self-energy at finite temperature. The diagrams we consider are
shown in Fig.~2 of the main text. The calculation presented here is performed for a single spin. At finite temperature, the asymptotic expression for the propagator reads
\begin{eqnarray}
G(\mathbf{z},\mathbf{z}';i\omega_n)\simeq e^{-i\chi } I(\mathbf{z}-\mathbf{z}';i\omega_n)
\hat{M}(\mathbf{z}-\mathbf{z}';\text{sgn}(n)),
\end{eqnarray}
where $i\omega_n$ is the fermionic Matsubara frequency. The function $\chi$ is defined by Eq. (\ref{chi}).
\subsection{Hartree and Fock diagrams}
In the coordinate space, the Hartree and Fock corrections to the self-energy are, respectively, given by the summation over bosonic Matsubara frequencies
\begin{eqnarray}
\Sigma_{\text{Hartree}}(\mathbf{x}_1,\mathbf{x}_2;i\omega_n)&\simeq& -T U^2
\sum_{i\nu_m} G(\mathbf{x}_1,\mathbf{x}_2,i\omega_n-i\nu_m) \Pi( \mathbf{x}_2-\mathbf{x}_1,i\nu_m),\\
\Sigma_\text{Fock}(\mathbf{x}_1,\mathbf{x}_2;i\omega_n)&\simeq& T^2U^2\sum_{i\nu_m,i\nu_l} G(\mathbf{x}_1,\mathbf{x}_2;i\omega_n-i\nu_m)G(\mathbf{x}_2,\mathbf{x}_1;i\omega_n-i\nu_m-i\nu_l)G(\mathbf{x}_1,\mathbf{x}_2;,i\omega_n-i\nu_l) \nonumber.
\end{eqnarray}
Here $U$ is the short-ranged interaction potential (touching potential) in momentum space, $\nu_m=2\pi T m$ and $\Pi( \mathbf{x}_2-\mathbf{x}_1,i\nu_m)$ is the polarization operator. Since the chiral structure of graphene suppresses the backscattering process, we focus on the zero-momentum species of polarization operator $\Pi_0$,
\begin{eqnarray}
\Pi_0(\mathbf{x_2},\mathbf{x_1};i\nu_m) &=& \frac{k_F^2}{2\pi^2 v_F^2} \frac{1}{ k_F r}
|\nu_m| e^{-|\nu_m| r/v_F}.
\end{eqnarray}
Here we introduced $r=| \mathbf{x_2}- \mathbf{x_1}|$. Further, we consider quasi-particles characterized by index $\nu$ ($\nu$ is characterized by the momentum for a free particle in the absence of the $B$-field and the Landau level index in the presence of the magnetic field). Then the self-energy is defined by
\begin{eqnarray}
\label{s5}
\Sigma(\nu,i\omega_n)=
\textbf{tr}\int d^2x_1 d^2x_2 A(\mathbf{x}_2,\mathbf{x}_1,\nu) \Sigma(\mathbf{x}_1,\mathbf{x}_2,i\omega_n) /\Big(V k_F/2\pi v_F\Big),
\end{eqnarray}
where $A$ is the spectral function of electrons in the absence of interaction and $V$ is the volume of the system. The spectral function is defined from by $A=-(2\pi i)^{-1} (G_R-G_A)$. Here $G_{R/A}$ are the retarded/advanced non-interacting Green functions. Taking an analytic continuation $i\omega_n \rightarrow \omega+i\delta$ (consider the case with $\omega>0$) and separating the on-shell singularity, we obtain
\begin{eqnarray}
\Sigma_{\text{Hartree}}(\omega+i\delta )\simeq i \Big(\frac{k_FU}{v_F}\Big)^2 \frac{T}{ 2\pi^2 E_F} \int \frac{e^{-\delta r/v_F} dr}{ r}
\Bigg[
\frac{\pi T} {\sinh^2(2\pi T r/v_F)}
- \nonumber \\ \frac{e^{i2\omega r/v_F }}{ \sinh(2\pi T r/v_F)} \Big( -i\omega + \frac{ \pi T }{ \tanh(2\pi T r/v_F)}\Big)
\Bigg], \nonumber
\end{eqnarray}
and
\begin{eqnarray}
\Sigma_\text{Fock}(\omega+i\delta)&\simeq & -i \Big(\frac{k_FU}{v_F}\Big)^2 \frac{T}{ 2\pi^2 E_F} \int dr \frac{ e^{-\delta r/v_F} \sin^2 \varphi(r)}{ r}
\Bigg[
\frac{\pi T } {\sinh^2(2\pi T r/v_F)}
- \nonumber\\
&& \frac{e^{i2 \omega r/v_F }}{ \sinh(2\pi T r/v_F)} \Big( -i\omega + \frac{ \pi T }{ \tanh(2\pi T r/v_F)}\Big)
\Bigg].
\end{eqnarray}
\subsection{$T/\omega\ll 1$}
For small $T/\omega\ll 1$, the
%(including the zero-temperature limit), 
$\Sigma_{\text{Hartree}}$ is simplified to
\begin{eqnarray}
\Sigma_{\text{Hartree}}(\omega+i\delta )\simeq i \Big(\frac{k_FU}{v_F}\Big)^2 \frac{1}{ 4\pi^3 } \int dr\frac{e^{-\delta r/v_F} dr}{ k_F r^2}
\Bigg[
\frac{v_F} { 2 r }
-e^{i2\omega r/v_F } \Big(- i\omega + \frac{ v_F}{ 2 r }\Big)
\Bigg],
\end{eqnarray}
and the $\Sigma_{\text{Fock}}$ becomes
\begin{eqnarray}
\Sigma_{\text{Fock}}(\omega+i\delta )\simeq -i \Big(\frac{k_FU}{v_F}\Big)^2 \frac{1}{ 4\pi^3 } \int dr\frac{e^{-\delta r/v_F} dr}{ k_F r^2}
\Bigg[
\frac{v_F} { 2 r }
-e^{i2\omega r/v_F } \Big(- i\omega + \frac{ v_F}{ 2 r }\Big)
\Bigg] \sin^2\varphi (r)\nonumber.
\end{eqnarray}
\begin{itemize}
	\item {\em Quasiparticle lifetime}. Here we calculate the imaginary part of the self-energy and find the expression for the lifetime. Defining dimensionless quantity $x=2\omega r/v_F$, we write $\Sigma_{\text{Hartree}}(\omega+i\delta )$ as
	\begin{eqnarray}
	\textbf{Im}\Sigma_{\text{Hartree}}(\omega+i\delta )\simeq \Big(\frac{k_FU}{v_F}\Big)^2 \frac{\omega^2}{ 2 \pi^3 E_F } \int^{+\infty}_{2\omega/E_F} dx
	\Big(
	\frac{1}{x^3} [
	1
	-\cos x
	]-\frac{1}{x^2} \sin x
	\Big).
	\end{eqnarray}
	The leading term in the integral $\int^{+\infty}_{2\omega/E_F} dx
	\Big(
	\frac{1}{x^3} [
	1
	-\cos x
	]-\frac{1}{x^2} \sin x
	\Big)$ is $2^{-1}\log(2\omega/E_F)$. Thus one obtains $
	\textbf{Im}\Sigma_{\text{Hartree}}(\omega+i\delta )\simeq \Big(\frac{k_FU}{v_F}\Big)^2 \frac{\omega^2}{ 4 \pi^3 E_F } \log(2\omega/E_F) $, which is the typical behavior of the lifetime of 2D Fermi liquids.
	The scattering rate $\delta=-\tau^{-1}_{\text{ee}}/2$ is estimated by the RPA diagrams which give the leading contribution to the lifetime of quasi-particles. Now we study the $B$-dependence of the lifetime. We consider the weak magnetic field such that the mean free path is smaller than the Lamour radius ($\delta>\omega_0$). Consider two cases.
	
	The first case is $T\ll \tau^{-1}_{\text{ee}}$. In this limit, the temperature is an irrelevant scale and the RPA lifetime dominates. We obtain the following equation fo the field-dependent self-energy term:
	\begin{eqnarray}
	\textbf{Im}\Sigma_{\text{Fock}}(\omega+i\delta )\simeq -\Big(\frac{k_FU}{v_F}\Big)^2 \frac{\omega_0^2}{16 \pi^3 E_F } \int^{+\infty}_{2\omega/E_F} \frac{e^{-\epsilon x} dx}{ x}
	\Big[
	1
	-\cos x
	\Big] .
	\end{eqnarray}
	Here $\epsilon=\delta/2\omega$. The leading term in the integral $\int^{+\infty}_{2\omega/E_F} \frac{e^{-\epsilon x} dx}{ x}
	\Big[1 -\cos x	\Big] $ giving $\log(2\omega\tau_{\text{ee}})$.
	
	The second case is $T\gg \tau^{-1}_{\text{ee}}$. In this limit, the temperature enters under the logarithm replacing the inverse lifetime there. Using the relation $\tau^{-1}=-2\textbf{Im} \Sigma$, the similar calculation yields the expression for lifetime given by Eq.~9 of the main-text.
	\item {\em Spectrum.} The real part of the self-energy gives a correction to the spectrum of quasi-particles. Thus here we compute $\textbf{Re}\Sigma_{\text{Hartree}}$ and $\textbf{Re}\Sigma_{\text{Fock}}$
	\begin{eqnarray}
	\textbf{Re}\Sigma_{\text{Hartree}}(\omega+i\delta )\simeq \Big(\frac{k_FU}{v_F}\Big)^2 \frac{1}{ 2\pi^3 } \frac{ \omega^2}{E_F} \int_{2\omega/E_F}^{+\infty} dx \Big[-\frac{ 1}{ x^2}
	\cos x + \frac{ 1}{ x^3}
	\sin x \Big].
	\end{eqnarray}
	The leading term in integral $\int_{2\omega/E_F}^{+\infty} dx \Big[-\frac{ 1}{ x^2}
	\cos x + \frac{ 1}{ x^3}
	\sin x \Big] $ is simply a constant, given by $\pi/4$. Thus this leads to $\textbf{Re}\Sigma_{\text{Hartree}}(\omega+i\delta )\simeq \Big(\frac{k_FU_0}{v_F}\Big)^2 \frac{1}{ 8\pi^2 } \frac{ \omega^2}{E_F} $
	and
	\begin{eqnarray}
	\textbf{Re}\Sigma_{\text{Fock}}(\omega+i\delta )\simeq \Big(\frac{k_FU}{v_F}\Big)^2 \frac{1}{ 32\pi^3 } \frac{\omega^2_0}{E_F}\int_{2\omega/E_F}^{\omega/\omega_0} dx e^{-\epsilon x}
	\Big( \cos x-\frac{1}{x}\sin x \Big).
	\nonumber
	\end{eqnarray}
	Then we extend the integral from $0$ to $\infty$, since $\omega / E_F \ll 1$ and  $\epsilon \omega/\omega_0\gg 1$. This yields
	\begin{eqnarray}
	\label{S18}
	\textbf{Re}\Sigma_{\text{Fock}}(\omega )\simeq -\Big(\frac{k_FU}{v_F}\Big)^2 \frac{1}{ (8\pi)^2 } \frac{\omega^2_0}{E_F}.
	\end{eqnarray}
	The expression is valid when $\tau_{\text{ee}}^{-1}(\omega) > \omega_0$. This introduces the correction to the quasi-particle spectrum in the low temperature limit, shown in Eq.~11. If one considers $\omega<0$, one will restore a $\text{sgn}(\omega)$ to Eq.~\ref{18}.
	
\end{itemize}

\subsection{$T/\omega\gg 1$}
In case of large $T/\omega \gg 1$, the temperature is a relevant scale.
\begin{itemize}
	\item {\em Lifetime}. Here we compute the imaginary part of the self-energy to find the quasiparticle lifetime. Defining dimensionless variable $x=\pi T r/v_F$, we write $\Sigma_{\text{Hartree}}(\omega)$ as
	\begin{eqnarray}
	\textbf{Im}\Sigma_{\text{Hartree}} \simeq -\Big(\frac{k_FU }{v_F}\Big)^2 \frac{T^2}{ 4 \pi E_F } \int^{+\infty}_{\pi T/E_F} \frac{dx}{x}
	\frac{1}{\cosh^2 x}.
	\end{eqnarray}
	The integral $\int^{+\infty}_{\pi T/E_F} \frac{dx}{x}
	\frac{1}{\cosh^2 x}$ is equal to $\log(E_F/\pi T )+c+O(T/E_F)$. Here $c$ is a constant. Thus we find $\textbf{Im}\Sigma_{\text{Hartree}}\simeq \Big(\frac{k_FU_0}{v_F}\Big)^2 \frac{T^2}{4 \pi E_F }\log( \pi T/E_F )$, a typical result for 2D Fermi-liquids. Meanwhile, one could write $\Sigma_{\text{Fock}}(\omega)$ as
	\begin{eqnarray}
	\textbf{Im}\Sigma_{\text{Fock}}& \simeq& \Big(\frac{k_FU }{v_F}\Big)^2 \frac{1}{ 4\pi^3 } \frac{\omega^2_0}{4E_F} \int_{\pi T/E_F}^{\infty} dx
	\frac{x}{ \cosh^2 x} \nonumber.
	\end{eqnarray}
	The integral $\int_{\pi T/E_F}^{\infty} dx
	\frac{x}{\cosh^2 x} = \log(2)+O(T^2/E_F^2)$. Thus $\textbf{Im}\Sigma_{\text{Fock}} \simeq \Big(\frac{k_FU_0}{v_F}\Big)^2 \frac{1}{ 4\pi^3 } \frac{\omega^2_0}{4E_F} \log(2)$. This gives the correction to the lifetime, written in Eq.~10 of the main-text.
	\item {\em Spectrum.} The real part of $\Sigma_{\text{Hartree}}$ is given by
	\begin{eqnarray}
	\textbf{Re}\Sigma_{\text{Hartree}}(\omega+i\delta )\simeq -\Big(\frac{k_FU  }{v_F}\Big)^2 \frac{\omega T}{ 2\pi^2 E_F} \int_{2\pi T /E_F}^{\infty} \frac{ dx}{ x}
	\frac{1 }{ \sinh x} \Big( 1 - \frac{ x}{ \tanh x}\Big).
	\nonumber
	\end{eqnarray}
	The integral $\int_{2\pi T /E_F}^{\infty} \frac{ dx}{ x}
	\Big[
	\frac{1 }{ \sinh x} \Big( 1 - \frac{ x}{ \tanh x}\Big)
	\Big]=-\log(2)+O(T/E_F)$. Thus we find that $\textbf{Re}\Sigma_{\text{Hartree}}(\omega+i\delta )\simeq \Big(\frac{k_FU_0}{v_F}\Big)^2 \frac{\omega T}{ 2\pi^2 E_F} \log(2)$. This is also a typical result for 2D Fermi liquids. Then we consider the Fock diagram and write $\textbf{Re}\Sigma_{\text{Fock}}$ as
	\begin{eqnarray}
	\textbf{Re}\Sigma_\text{Fock}(\omega+i\delta)&\simeq & \Big(\frac{k_FU }{v_F}\Big)^2 \frac{\omega \omega_0^2 }{ 32\pi^4 E_F T} \int_{2\pi T/E_F}^{2\pi T/\omega_0} dx \times x
	\frac{ 1}{ \sinh x} \Big( 1 -\frac{x }{ \tanh x}\Big).
	\end{eqnarray}
	Since $T/E_F\ll 1$ and $T/\omega\gg 1$, we can replace the lower bound of the integral domain by $0$ and the upper bound by $\infty$. Making use of the integral $\int_{0}^{+\infty} dx x
	\frac{ 1}{ \sinh x} \Big( 1 -\frac{x }{ \tanh x}\Big) =-\pi^2/4$, we obtain at $T\gg \omega,\omega_0$:
	\begin{eqnarray}
	\label{S27}
	\textbf{Re}\Sigma_\text{Fock}(\omega+i\delta)&\simeq & - \Big(\frac{k_FU  }{v_F}\Big)^2 \frac{ \omega }{ 2 (8\pi)^2 } \frac{\omega_0^2 }{E^2_F} \times \frac{E_F}{T} .
	\end{eqnarray}
	Equations ~\ref{S18} and ~\ref{S27} result in Eq. (11) of the main text. 
	
	The velocity of quasi-particle might be obtained by Fourier transforming Eq.~\ref{chi} by $ (2\pi)^{-1}\int  d\omega G(\omega) \exp(i\omega t)$. One obtains a delta-funtion $\delta(r-v_F t)$ after the transformation. This indicates that the velocity of quasi-particle is $v_F$ in the semi-classical regime, even in the presence magnetic field. In the presence of interaction, the spectrum of quasi-particle is corrected. Thus the propagator in Eq.~\ref{chi} obtains interaction correction. Now the exponential part in $G(\omega)$ becomes $\exp i \Big(\omega-	\textbf{Re}\Sigma(\omega) \Big) r/v_F $, where $\Sigma$ is the self-energy correction, containing both Hartree and Fock corrections. Then the Fourier transformation leads a delta function, $\delta \left(\left(1-\partial_\omega \textbf{Re}\Sigma|_{\omega=0}\right)r-v_Ft \right)$. Thus the velocity got renormalzied by the interaction to be $v_F/\left(1-\partial_\omega \textbf{Re}\Sigma|_{\omega=0}\right)\simeq v_F(1+\partial_\omega \textbf{Re} \Sigma|_{\omega=0} )$. Since the field-dependence correction comes from Eq.~\ref{S27}, one then uses Eq.~\ref{S27} to obtain the field-dependent correction to effective velocity and then recovers Eq.~12 in the main text.
	
	%renormalization is also from the self-energy in Eq.~\ref{S27}. In the limit $T\gg \omega$, the energy quasi-particle obtains a correction $\epsilon(\omega,B)-\epsilon(\omega,0)=\textbf{Re}\Sigma_\text{Fock}(\omega+i\delta)$.
\end{itemize}
%%%%%%%%%%%
%%%%%%%%%%%
%%%%%%%%%%%
\section{Derivation of the specific heat}
\begin{figure}
	\centering
	\includegraphics[scale=0.25]{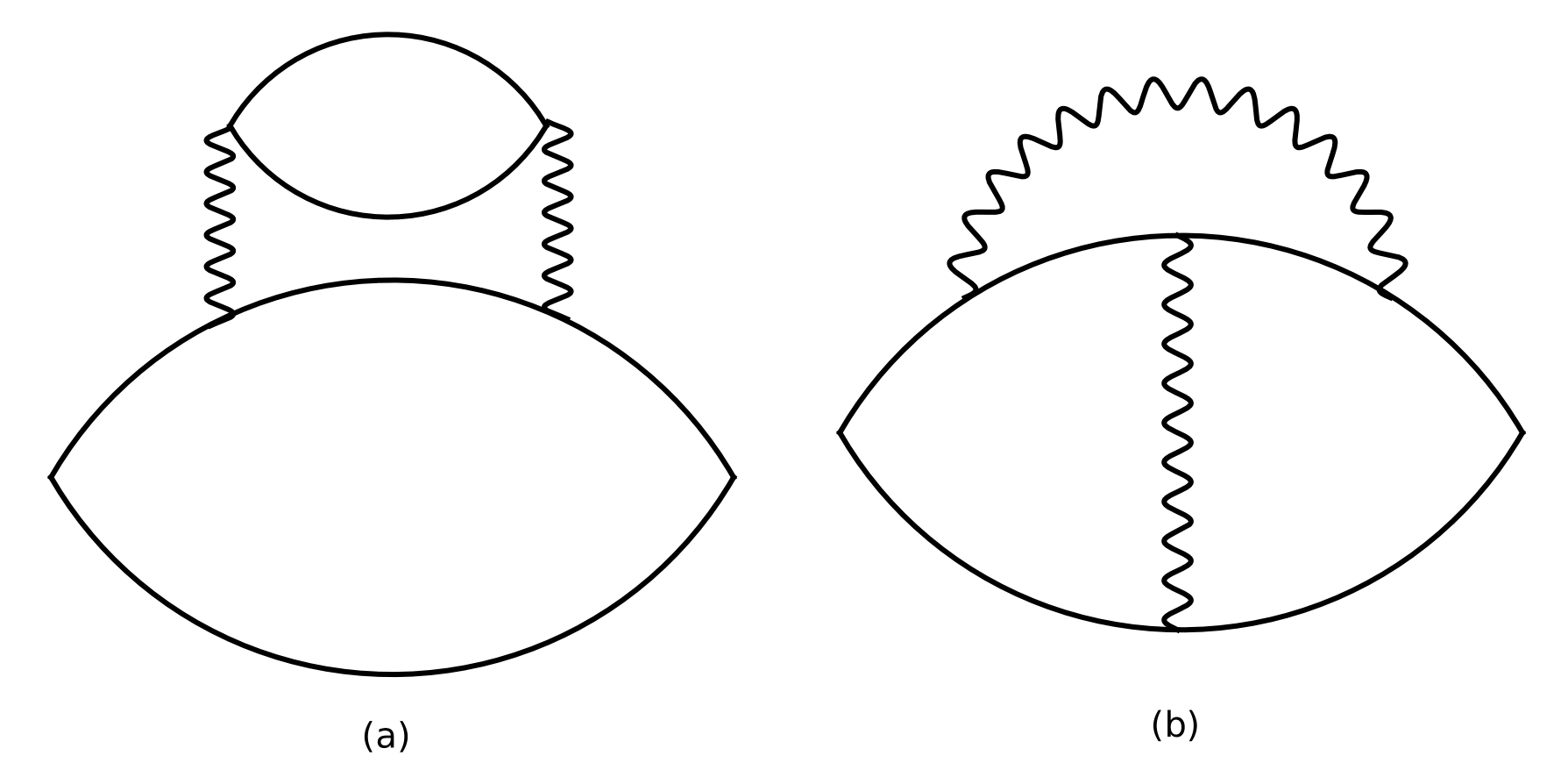}
	\caption{The Feynman diagrams representing the corrections to thermodynamic potential. (a) Hartree diagram. (b) Fock diagram.}
	\label{d}
\end{figure}
According to the linked cluster theorem, the thermodynamical potential is reduced to the computation of closed Feynman diagrams. We employ perturbation theory in interactions and consider two leading diagrams depicted in Fig.~(\ref{d}). The calculation presented here is for a single spin.
The Hartree term, given by the diagram Fig.~(\ref{d})a, gives the following contribution to the potential
\begin{eqnarray}
\Omega_a=-T \sum_{\nu_m}\Pi(i\nu_m)\Pi(i\nu_m).
\end{eqnarray}
Here $\nu_m=2\pi m T$. The Fock term, depicted in Fig.~(\ref{d})b, gives
\begin{eqnarray}
\label{s31}
\Omega_b= T^3 \textbf{tr}\sum_{m,n,l} G(i\omega_m) G(i\omega_m-\nu_l)G(i\omega_m-\nu_l-i\nu_n) G(i\omega_m-i\nu_n).
\end{eqnarray}
\subsection{Hartree diagram}
Due to the chiral structure, the leading contribution to $\Omega_a$ is from the zero-momentum transfer. Upon summing over $\nu_m$ and defining dimensionless variable $x=2\pi T r/v_F$, one obtains
\begin{eqnarray}
\Omega_a &=&-v U^2 T^3 \frac{k_F^2}{ \pi v_F^4} \int_{2\pi T/E_F}^{\infty} \frac{dx}{x} \frac{1}{( \sinh x)^{2}\tanh x}.
\end{eqnarray}
The integral $\int_{2\pi T/E_F}^{\infty} \frac{dx}{x} \frac{1}{( \sinh x)^{2}\tanh x}=c_0(2\pi T/E_F)^{-3}-c_1+O(T/E_F)$. Here $c_0$ and $c_1$ are two constants. Importantly, one may write $c_1$ as a generic integral
$
c_1=\int_{0}^{\infty} \frac{dx}{x} \Big[
\frac{1}{x^3} -\frac{1}{( \sinh x)^{2}\tanh x}
\Big]
$
Numerically, it is approximately given by $c_1\approx 0.121727$.
Then we consider the specific heat per unit volume
\begin{eqnarray}
C^a_v=\frac{1}{v} \frac{\partial \Omega_a}{\partial T}\Big|_v .
\end{eqnarray}
Obviously, the leading contribution comes from the $\propto c_1$ part. Thus we obtain
\begin{eqnarray}
C^a_v=3c \Big(\frac{k_F U }{v_F}\Big)^2 \frac{T^2}{ \pi v_F^2}.
\end{eqnarray}
This $T^2$ behavior is a typical property of 2D Fermi-liquids.
\subsection{Fock diagram}
Upon the summation over $m,n,l$ in Eq.~(\ref{s31}), we obtain the following integral expression
\begin{eqnarray}
\label{31}
\Omega_b= \frac{vT}{ (2\pi)^3} \Big( \frac{k_F U }{ v _F} \Big)^2 \frac{\omega_0^2}{2v_F^2} \int^{+\infty}_{2\pi T/E_F} zdz \frac{1}{ \sinh^2z \tanh z}.
\end{eqnarray}
The integral
$ \int^{+\infty}_{2\pi T/E_F} zdz \frac{1}{ \sinh^2 z \tanh z}=a_0 \frac{E_F}{T}-\frac{1}{2}+O(T^3/E^3_F)$. Here $a_0$ is a constant. Thus its contribution to specific heat, $C^b_v$, now reads
\begin{eqnarray}
C^b_v\simeq -\frac{ 1}{ (2\pi)^3} \Big( \frac{k_F U }{ v _F} \Big)^2 \frac{\omega_0^2}{ 4v^2_F}.
\end{eqnarray}
This is a constant correction to specific heat, which only depends on the magnetic field $B$, $v_F$, $k_F$ and the interaction strength $U_0$.

\section{Zeeman effect}
In this section, we argue that the Zeeman effect gives subleading corrections to thermodynamics compared to the (pseudo)-spin phase. The Zeeman effect originate from the coupling
$-\frac{1}{2}\mu_B B1_4\otimes \hat\alpha_z.
$
Here $\mu_B$ is the Bohr magneton and $\hat\alpha_z$ is the Pauli matrix acting on the (real-)spin of electron.
Since energy is splitted by the Zeeman effect, each spin's Fermi momentum is different. Namely,
\begin{eqnarray}
k^s_F=k_F+\frac{s}{2}\frac{\mu_B B}{v_F},
\end{eqnarray}
where $k_F$ is defined by $\epsilon_F/v_F$.  We find that the propagator is modified to be
$
G_{K,s}(\mathbf{r},\delta w) \approx \frac{I_s}{2}  \Big(1+(\text{sgn}(\omega) +\frac{i}{2k^s_Fr})\hat{r}\cdot \hat{\sigma}\Big)
$.
Two observations: (1) The chiraly feature is not influenced by the Zeeman effect. (2) $k^s_F$ enters the oscillatory function. This is the key point to consider the Zeeman effect.
In fact, it affects the thermodynamic potential in the following way
\begin{eqnarray}
\Omega^z_a=-T U^2_{2k_F}  \sum_{\nu_m} \int d^2r\Pi_+(i\nu_m,r)\Pi_-(i\nu_m,r).
\end{eqnarray}
Here $\Omega^z_a$ is the correction to the thermodynamic potential by Zeeman effect and $\Pi_{\pm}$ is $2k_F$ polarization operator for spin up and down respectively. Here $\Pi_{2k_F}^s$ is the $2k_F$ polarization operator for the spin-up/down $(s=+/-)$ electron. However, in Graphene, the $2k_F$-scattering is suppressed by chirality. In fact, $\Pi_{2k_F}^+(r) \Pi_{2k_F}^-(r)$ carries one extra decaying factor $(k_Fr)^{-2}$. Since the characteristic scale for $r$ is the thermal length $v_F/T$, the decaying factor then translates into a small factor, $T^2/E_F^2$.

In the Zeeman effect, the fundamental energy scale is
$
\omega_{z}=\mu_B B
$.
Set $B=x (T)$ and then $\omega_z$ given by
\begin{eqnarray}
\omega_z=9.27x \times 10^{-24}  J.
\end{eqnarray}
The effective cyclotron-frequency is given by $\omega_0=\hbar v_F/R_L$. If $k^{-1}_F\simeq 1.65$ nm, the Larmor radius given by
$
R_L\simeq \frac{400}{x} nm=4\times 10^{-7}/x (m)
$.
For simplicity, we take $v_F=10^{6}m/s$. Recall $\hbar\simeq 10^{-34} J\cdot s$. Then the energy scale $\omega_0$ is estimated by
\begin{eqnarray}
\omega_0=  \frac{x}{4}\times 10^{-21} J.
\end{eqnarray}
It is easy to see that $\omega_0\gg \omega_z$.

From smallness of $T^2/E_F^2$ and $\omega_z/\omega_0$, one observes that the Zeeman effect gives parametrically smaller corrections compared to the effect coming from the pseudospin-dependent orbital phase.

\section{ Effect of curving of the electron trajectory}
In this section, we consider the effect of curving of the electron trajectory  and argue that it gives a sub-leading 
correction to the thermodynamic characteristics of the interacting electron gas. 
% compared to the effect in the main-text.
% The curved trajectory effect is reflected
Curving of the trajectory manifests itself in the two-loop diagram in the RPA sequence. Now we consider the correction to DOS and find that it is proportional to the following expression
\begin{eqnarray}
\label{63}
P(\mathbf{r},\omega) =U^2_{2k_F} u_0^2\text{tr}\int d^2 r_1 d^2 r_2 G(\mathbf{r},\mathbf{r}_1;\omega)G(\mathbf{r}_1,\mathbf{r}_2;\omega)G(\mathbf{r}_2,\mathbf{r};\omega)\Pi(\mathbf{r}_1;0)\Pi(\mathbf{r}_2;0),
\end{eqnarray}
where $U_{2k_F}$ is the $2k_F$ component of interaction and $u_0$ is the strength of the 
%diagonal
point-like impurity.
Since we focus on the effect trajectory, we only preserve the $p_0^3r^3$-corrections and ignore the spin-dependent phase correction in $P$. We call this part as $P_0$, given by the following integral expression
\begin{eqnarray}
P_0= &&-\frac{\partial_\omega}{2}\text{tr} \int d^2 r_1 d^2 r_2    \frac{1}{2\pi v_F^2 |\mathbf{r}_1-\mathbf{r}_2|^2  }   \exp  2i \left((k_F+\frac{\omega}{v_F})|\mathbf{r}_1-\mathbf{r}_2|- \frac{  p_0^3|\mathbf{r}_1-\mathbf{r}_2|^3}{24}  \right)    \nonumber\\
&&\times  \frac{1}{ 4\pi^2 v_F r_1^3 }   \cos(2k_Fr_1-\frac{p^3_0 r_1^3}{12}) \times    \frac{1}{ 4\pi^2 v_F r_2^3 }   \cos(2k_Fr_2-\frac{p^3_0 r_2^3}{12})  U^2_{2k_F} u_0^2.
\end{eqnarray}
In the exponential, the main oscillating term is $2k_F|\mathbf{r}_1-\mathbf{r}_2|\pm 2k_Fr_1\pm 2k_F r_2$. We only need the slowly oscillatory piece. Namely, we want $2k_F|\mathbf{r}_1-\mathbf{r}_2|\pm 2k_Fr_1\pm 2k_F r_2\simeq 0$. This demands that $\mathbf{0}, \mathbf{r}_1, \mathbf{r}_2$ are aligned in a straight line. For simplicity, we take $\mathbf{r}_1/r_1\simeq -\mathbf{r}_2/r_2$.  Use the expression
$
|\mathbf{r}_1-\mathbf{r}_2|=|r_1+r_2|\sqrt{1-\frac{2r_1r_2}{|r_1+r_2|} (1+\cos  \theta  ) }
$
where $\theta$ is the angle bewteen $\mathbf{r}_1$ and $\mathbf{r}_2$. If $\theta\simeq \pi$, then $|\mathbf{r}_1-\mathbf{r}_2|\simeq r_1+r_2+\frac{ r_1r_2}{2|r_1+r_2|}( \theta- \pi)^2 $. Then the integral over $\theta$ introduces extra factor, shown as below  \begin{eqnarray}
\int_{\theta \sim \pi} d \theta \exp (ik_F \frac{r_1r_2}{|r_1+r_2|} (\theta-\pi)^2)\approx  e^{i\pi/4}\sqrt{\frac{\pi |r_1+r_2|}{k_F r_1 r_2}}.
\end{eqnarray}
%Thus the $P_0$ contain the following integral\begin{eqnarray} P_0^a&\equiv&-i \int_{r_1,r_2>k_F^{-1}} dr_1 d r_2    \frac{1}{  v_F^3 (r_1+r_2)   }   \exp  2i \left( \frac{\omega}{v_F}(r_1+r_2)-  \frac{  p_0^3 \left[ r_1^2r_2+r_1 r_2^2\right]}{8}  \right)    \nonumber\\&&\times  \frac{1}{ 4\pi^2 v_F r_1^2 }     \times    \frac{1}{ 4\pi^2 v_F r_2^2 } \times  e^{i\pi/4}\sqrt{\frac{\pi ( r_1+r_2) }{k_F r_1 r_2}} U^2_{2k_F} u_0^2\end{eqnarray}Here we used $r_1^3+r_2^3-(r_1+r_2)^3=-3r_1^2r_2-3r_1 r^2_2$.
We only trace the field-dependent correction, by considering $\delta P_0^a(B)=P_0^a(B)-P_0^a(0)$.
Further we do the following variable change: $x=p_0 r_1$ and $y=p_0 r_2$. Here $x$ and $y$ are both dimensionless. Then $\delta P_0^a$ is given by $U^2_{2k_F} u_0^2 v_F^5 k_F^{-1/2}p_0^{-7/2}F_\Gamma(\omega/\epsilon_0)$ and $F_\Gamma$ is given by
\begin{eqnarray}
F_\Gamma&\equiv&-i  \int_{x,y>\Gamma} dx  d y    \left[
\exp  2i \left( \frac{\omega}{\epsilon_0}(x+y)-  \frac{   \left[ x^2y+x y^2\right]}{8}  \right) -  \exp  2i \left( \frac{\omega}{\epsilon_0}(x+y)  \right)
\right]   \nonumber\\
&&\times  \frac{1}{ 4\pi^2  x^2 }     \times    \frac{1}{ 4\pi^2  y^2 } \times  e^{i\pi/4}\sqrt{\frac{\pi  }{  xy (x+y)}} \label{70},
\end{eqnarray}
where $\Gamma\ll 1$ is small cut-off with order of $p_0/k_F$ and $\epsilon_0=v_F p_0$. Numerically, we trace the $\Gamma$-independent term, namely $F(\omega/\epsilon_0)$, in Eq.~\ref{70} and find that $F(\omega/\epsilon_0)$ is smooth function of $\omega/\epsilon$ and has the amplitude that is smaller than $1$. Thus, we conclude that the curved trajectory give a correction to the DOS with a coefficient, $\propto B^{7/2}$. Since we consider a weak magnetic field,  the $B^{7/2}$ is obviously the high order perturbation relative to $B^2$. Further, $F(\omega/\epsilon_0)$ is smooth and smaller than $1$, while $E^2_F/\omega^2$ is large. Thus the effect of curved trajectory gives the sub-leading corrections.
%\section{Other diagrams}The lower order perturbations, including one impurity scattering, two impurity scatterings, one impurity scattering and one interaction line, are giving highly oscillating terms to local tunneling DOS. So these diagrams give negligible contributions to DOS.Higher order perturbations? Two interaction lines and two impurity scatters includes (1) two loop corrections (2) involved small momentum transfer with two Fock self energy term.  (...)

\end{document}